\documentclass[11pt]{article}

\usepackage{hyperref}
\usepackage{dsfont}
\usepackage{bm}
\usepackage{float}
\usepackage{cases}
\usepackage{subfigure}
\usepackage{enumitem}
\usepackage{multirow}
\usepackage{amssymb, amsthm}
\usepackage{graphicx}
\usepackage[round]{natbib}
\usepackage{todonotes}
\usepackage{fancyhdr}
\usepackage{phaistos}

\usepackage{setspace}
\usepackage{fancyhdr}
\title{Analysing ecological dynamics with relational event models: the case of biological invasions}

\date{}

\author{Ruta Juozaitiene\\[2pt]
	\textit{VU Institute of Data Science and Digital Technologies, Vilnius, Lithuania;}\\ \textit{Vytautas Magnus University, Kaunas, Lithuania}\\[6pt]
	Hanno Seebens\\[2pt]
	\textit{Senckenberg Biodiversity and Climate Research Centre, Frankfurt, Germany}\\[6pt]
	Guillaume Latombe \\[2pt]
	\textit{Institute of Evolutionary Biology, University of Edinburgh, UK}\\[6pt]
	Franz Essl\\[2pt]
	\textit{Bioinvasions. Global Change. Macroecology Group,}\\
	\textit{Department of Botany and Biodiversity Research, 
		University of Vienna, Austria}\\[6pt]	
	Ernst C. Wit\footnote{To whom correspondence should be addressed: \underline{wite@usi.ch}} \\[2pt]
	\textit{Institute of Computing, Universit\`a della Svizzera italiana, Lugano, Switzerland}
}

\begin{document}

\maketitle

\begin{abstract}
	\noindent	\textbf{Aims}\\
		Spatio-temporal processes play a key role in ecology, from genes to large-scale macroecological and biogeographical processes. Existing methods studying such spatio-temporally structured data either simplify the dynamic structure or the complex interactions of ecological drivers. The aim of this paper is to present a generic method for ecological research that allows analysing spatio-temporal patterns of biological processes at large spatial scales by including the time-varying variables that drive these dynamics.
		
	\noindent	\textbf{Location}\\
		Global analysis at the level of 272 regions. 
		
	\noindent	\textbf{Methods}\\
		We introduce a method called relational event modelling (REM). REM relies on temporal interaction dynamics, that encode sequences of relational events connecting a sender node to a recipient node at a specific point in time. We apply REM to the spread of alien species around the globe between 1880 and 2005, following accidental or deliberate introductions into geographical regions outside of their native range. In this context, a relational event represents the new occurrence of an alien species given its former distribution. 
		
	\noindent	\textbf{Results}\\
		The application of relational event models to the first reported invasions of 4835 established alien species outside of their native ranges from four major taxonomic groups enables us to unravel the main drivers of the dynamics of the spread of invasive alien species. Combining the alien species first records data with other spatio-temporal information enables us to discover which factors have been responsible for the spread of species across the globe. Besides the usual drivers of species invasions, such as trade, land use and climatic conditions, we also find evidence for species-interconnectedness in alien species spread.
		
	\noindent	\textbf{Conclusions}\\
		Relational event models offer the capacity to account for the temporal sequences of ecological events such as biological invasions and to investigate how relationships between these events and potential drivers change over time.

\end{abstract}

\section{Introduction}

The number of alien species has increased continuously over recent centuries \citep{Seebens2017} with far-reaching consequences on the biogeography of species, ecosystem functioning and native species \citep{pyvsek2020scientists}. The introduction of species beyond their native ranges is attributable to human mediated activities that have, deliberately or accidentally, enabled them to overcome fundamental geographical barriers. Rapid range expansion of an invasive species is a primary threat to global biodiversity and ecosystems. A better understanding of biological invasions may help identify potential strategies for preventing and mitigating invasion impacts. However, the dynamics of such inter-regional spread can be complex, varying in time, space and among species. In addition, dynamics are usually driven by multiple factors simultaneously, involving exogenous ecological and socio-economic factors, such as transport hubs \citep{floerl2009importance}, habitat suitability, climatic conditions or land use \citep{pyvsek2010disentangling,dyer2017global,Essl2019}. Given the complex interplay of natural and anthropogenic spatio-temporal factors, it is crucial to study the effects of multiple factors affecting alien species spread simultaneously. 

Existing statistical approaches aiming to predict the occurrence and spread of alien species, ranging from correlative \citep[e.g.][]{elith2009species} to dynamic modelling tools \citep{dullinger2009niche,chapman2016modelling,paini2012modelling}, are usually constrained to individual species and disregard the timing of an invasion in a particular region, ignoring potential temporal changes in the magnitude and even direction of drivers on invasions. In particular, species distribution models (SDMs) are widely used in ecology to estimate species geographic ranges and to understand the drivers of species distributions. However, these models are criticized for lacking important mechanisms (such as species dispersal, establishment, and biotic interactions) and independent validation, among other things \citep{araujo2012uses}. It is argued that the ability of SDMs to predict independent data is higher when considering single-species versus multi-species studies. A detailed literature review revealed that 77\% of studies focusing on a single species found that these models successfully predicted occurrence, whereas 23\% of studies incorporating more than one species found unambiguous support for the SDMs prediction abilities \citep{a2022species}. Moreover, of the vast majority previous research focused only on static SDMs, which do not account for temporal variation of predictor and response variables \citep{bellard2016major,ni2021botanic}. Other approaches suggest analysing residuals of a model with a single predictor to remove first the confounding effect, such as sampling effort \citep{sofaer2017accounting,bonnamour2021insect} or region area \citep{brown2021plant}. Several attempts to account for temporal variation in the drivers of alien species introduction performed separate analyses for each period of interest. For instance, multiple linear regression models were developed to compare relationships between the number of alien species and explanatory variables across different time periods \citep{dyer2017global,brown2021plant}. However, each subset consisted of aggregated data representing the overall period for each region. In a recent study, global temporal dynamics of alien species invasions were analysed using the annual data of trade openness and the number of invasions per year \citep{bonnamour2021insect}. In this case, the analysis of relationship between these two time-varying quantities was based on Pearson's product moment correlation. Thus, this approach does not allow to simultaneously quantify the effect size of multiple drivers and ignores possible confounding factors, that are very common when performing univariate modelling. 

This paper aims to describe a generative event history model that captures the time-varying dynamics of alien species introductions and spread by integrating potentially relevant factors. So-called relational event model was first developed in the field of social network analysis \citep{butts2008a, perry2013point,Juozaitiene2022}, before being applied in multiple other fields, for example to interactions among cattle \citep{patison2015time}, to transfers of patients between hospitals \citep{vu2017relational} and to political interactions \citep{dubois2010modeling}. A relational event describes a time-stamped interaction, such as the appearance of a species in a new location. A relational event model thus analyses sequences of relational events following their observed temporal order. 
Ecological invasion processes have crucial relational event characteristics: an invasion is a temporal event whereby a species becomes associated with a certain new region. The methodological advantage of using relational event models is the fact that they are able to describe the underlying general interaction patterns, can deal with time-varying variables and time-varying effects, and have a suit of diagnostic tools to test the suitability of the models on actual data. 


First, we describe the data of invasion events of four taxonomic groups (i.e. mammals, birds, insects, vascular plants) from the Alien Species First Record Database (FirstRecords), the most exhaustive source of first records of species in regions of the world \citep{Seebens2017}. Then, we introduce the relational event model (REM) for studying the factors driving the spread of alien species across large spatial (i.e. globally) and temporal (i.e. 1880-2005) scales. We analyse effects of various drivers on the timing of first records of alien species. Based on these results, we provide a synthesis of the advantages and caveats of using REMs for studying ecological invasions. We conclude our work with general ideas for REM application of analysing temporal sequences of ecological relational events in various ecological contexts.

\section{Methods}
\label{sec:materialmethods}

In this section, we first describe the data that are used in this study. Then we give an overview of the relational event model for ecological data and describe how it can be applied to the alien species invasion process.

\subsection{Data}\label{sec:data}

This study is based on the Alien Species First Records database \citep{Seebens2017}, which contains the years of the first records of established alien species across taxonomic groups in regions worldwide (i.e. countries, and sub-national regions such as islands, archipelagos and federal states). Established alien species refer to species introduced, accidentally or intentionally, outside of their natural geographic range and that have established a viable population there \citep{blackburn2011proposed}. The data contains more than 47,000 records of 16,922 established alien species across 275 regions worldwide. Most data are from plants (48\%), followed by insects (26\%) and birds (6\%), whereas mammals are present in 3\% of the records. Regions usually correspond to countries or islands having a particularly large number of samples. In this manuscript, we use first records of four major land-based taxonomic groups, namely vascular plants, insects, birds and mammals, during the period of from 1880 until 2005, which represent those taxonomic groups and times with most available information. The reason for the cut-off year of 2005 is data availability of first records. In previous studies \citep{Seebens2017, Seebens2018}, it has been found that the first records after 2005 are distinctly affected by a lag in reporting. It takes time until a species has been recognized in a new environment, assessed as being an alien species with established (i.e., reproducing) populations and the new record has been published. The estimate from previous studies is that this process takes on average 10-15 years, which is the reason why more recent first records cannot be reliably interpreted at a global scale. The selected taxonomic groups cover 81\% of the data contained in the Alien Species First Record Database. 

We consider a range of explanatory variables to characterise the spatial locations, climatic conditions, political relationships such as former colonial ties among countries and various socio-economic variables during that period (see Table~\ref{tab:data}). These variables have been shown to be of importance for explaining the current distribution of alien species \citep[e.g.][]{pyvsek2010disentangling,dawson2017global,Essl2019}. 

Land cover types have been shown to be associated with different levels of alien species presence \citep{chytry2009european,Polce2011,Ortiz2020}. Here, information on land cover  is used to characterise the potential suitability of regions for the establishment of alien species and the degree of anthropogenic disturbance, which facilitates establishments. Land cover is described by  the proportion of cropland, pasture and urban area; data were obtained from the Harmonized Global Land Use data set, V1 \citep{Hurtt2011}. The grid-based information has been averaged for the regions and years (1971-2012) considered in this study. As land cover data were provided on a decadal basis, while the analysis was run on an annual scale, land cover variables were interpolated from a 10-year resolution to obtain an annual resolution.

It is expected that favourable climate conditions facilitate the spread of alien species and unsuitable environments with stressful conditions often lead to a lower establishment success \citep{Finch2021}. To account for habitat suitability we use the air temperature at 2 meters above the ground as the near-surface air temperature. This variable is used to represent climate conditions, using simulated annual values covering the period from 1871 to 2012 \citep{Watanabe2011}. Annual near-surface air temperature values were averaged for each region and year between 1871 and 2012.

International trade has been identified as a major driver of alien species introductions (via increased propagule and colonization pressures), and therefore increases the probability of species to become established \citep{westphal2008,chapman2017global,Blackburn2020,bagnara2022simulating}. International trade values were extracted from the Correlates of War project \citep{Barbieri2009}. The dataset includes information on trade flows between countries from 1871 to 2009. The dyadic trade dataset values represent imports from one country to another in current U.S. dollars. The dataset has missing values because trade flow is unknown for some countries within certain time periods and some countries are not included in the dataset. To address this issue, missing values that appear at the beginning of the observational period were replaced with zeros in those time series that start at zero value. In other cases, i.e., for time series for which the first recorded trade value between countries $i$ and $j$ is not equal to zero, we extrapolated missing values using the log-linear model, $\log(\mbox{trade}(i,j,t)+1) = \alpha_{ij} +\beta_{ij} t$. This model assumes a linear relationship between the logarithm of trade volume between countries $i$ and $j$ at time $t$ and the time point $t$. In principle, any extrapolation model could be used for this purpose. However, given the nature of economic growth, an exponential growth model is appropriate and therefore was used to determine the missing values.

A substantial part of the most important trade routes has formed since colonial times. It is expected that former colonial ties facilitated the spread and establishment of alien species \citep{yang2021global,lenzner2022naturalized}. Colonial ties were therefore included in this analysis using the Colonial Dates Dataset (COLDAT) \citep{COLDAT}, which provides information on the reach and duration of European colonial empires. Each country is uniquely characterised by the colonial empire it belonged to or classified as independent, in case there were no colonial ties.

Previous studies revealed that observed invasion dynamics are highly correlated with variations in sampling effort and, thus, it is important to address this bias before reaching conclusions regarding temporal trends of biological invasions \citep{bonnamour2021insect,sofaer2017accounting}. To account for variations in sampling effort, we used the number of species present in a particular region in 1880 as a proxy. The number of present species was calculated by summing the number of First Records up until 1880 and the number of native species. This variable was used as an approximate measure of sampling effort under the assumption that there was no saturation of the pool of native species yet to be recorded in each country at that time and that sampling effort is constant over time. To acquire information on native ranges for all species considered in our analyses, we combined data from various sources. For mammals and birds, native ranges were extracted as range maps from the IUCN range maps (\url{https://www.iucnredlist.org}, accessed 08.07.2016) and BirdLife (\url{http://datazone.birdlife.org/species/requestdis}, accessed 01.11.2016), respectively. A species was then assigned to a region listed in the first record database if the range map intersected with the region. For vascular plants and insects, native ranges were extracted from \citep{van2019global} and the CABI Invasive Species Compendium (\url{https://www.cabi.org}, accessed 15.07.2016), respectively. Region names were harmonised to match the regions provided in the first record database. 

\begin{figure}[h]
	\centering
	\includegraphics[width=0.7\linewidth]{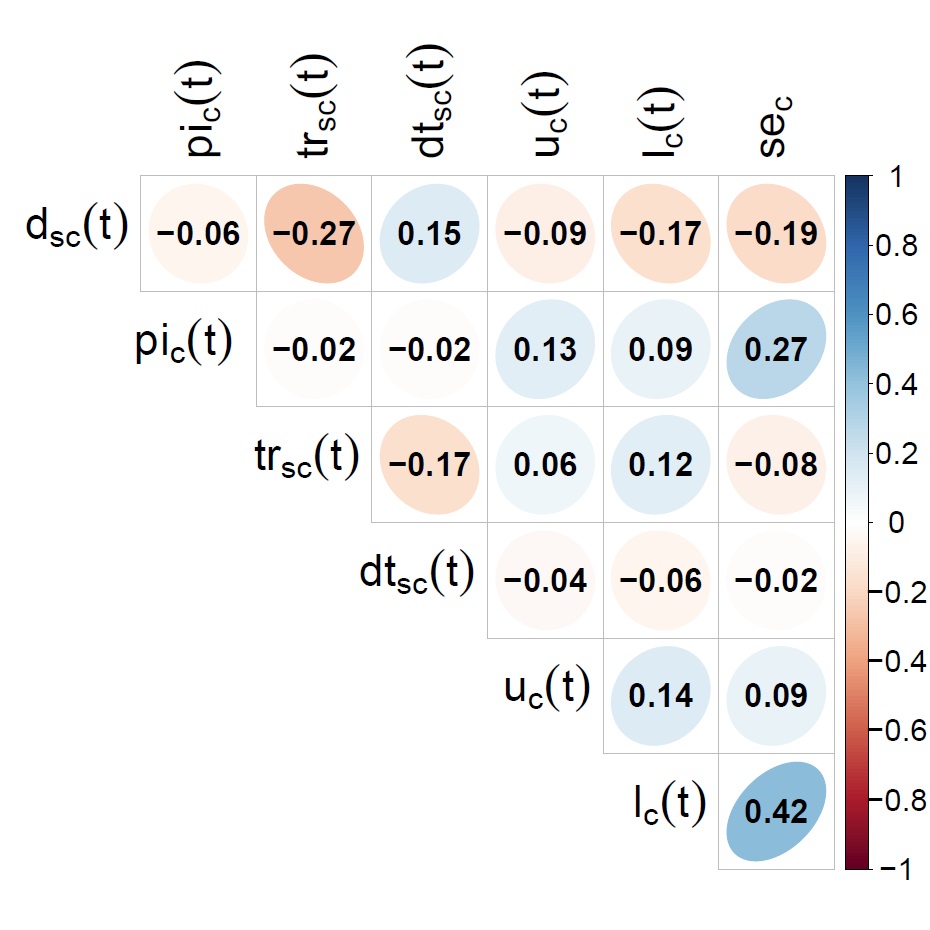}
	\caption{A heatmap representing correlations between the explanatory variables in the model. As can be seen, multicollinearity between the variables is low, suggesting that the estimation is stable.}
	\label{fig:corr}
\end{figure}

To confirm independence between explanatory variables, we used a correlation heatmap indicating correlations among all variables (Figure \ref{fig:corr}). Coefficient values were small with the highest observed correlation equal to 0.42. Therefore we conclude that multicollinearity does not cause a problem in the fitting of the model. 

\begin{figure}[h]
	\centering
	\includegraphics[width=0.7\linewidth]{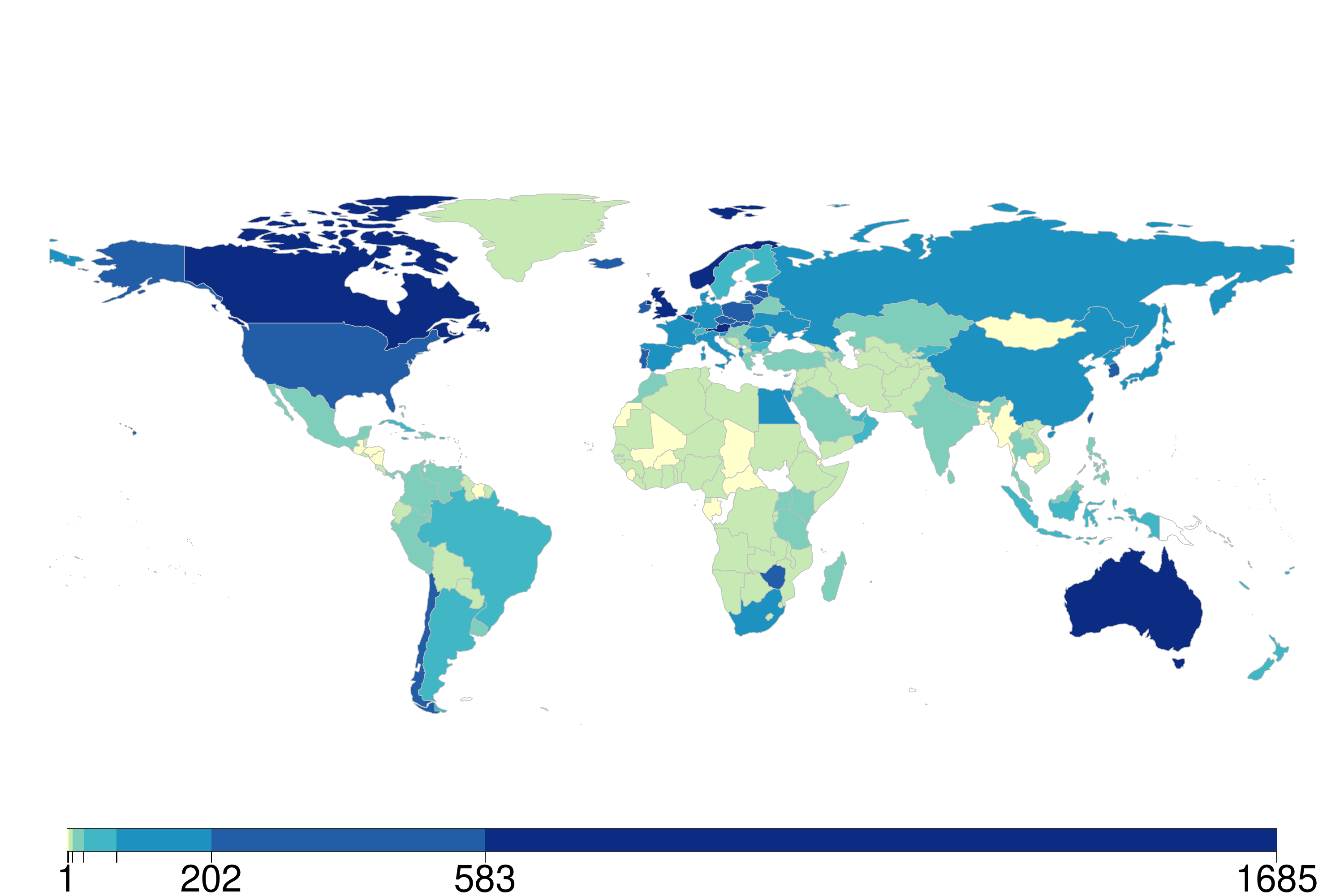}
	\caption{Global map of the number of invasion events per country. The total number of analysed invasion events is $n=16403$. Australia is the country with most first records ($n=1685$). }
	\label{fig:num_inv}
\end{figure}

\begin{figure}[h]
	\centering
	\begin{tabular}{cc}
		\includegraphics[width=0.5\linewidth]{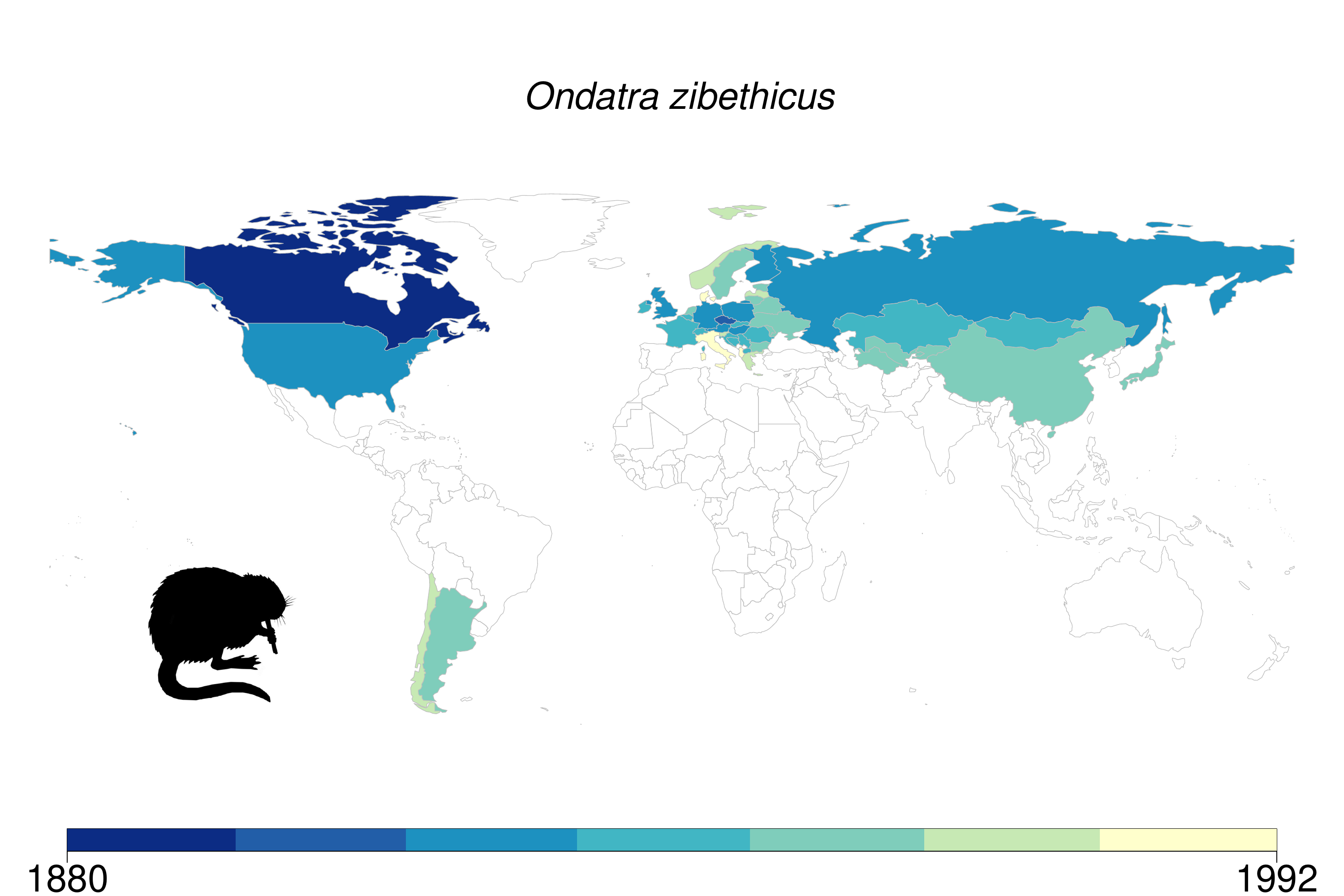}
		&
		\includegraphics[width=0.5\linewidth]{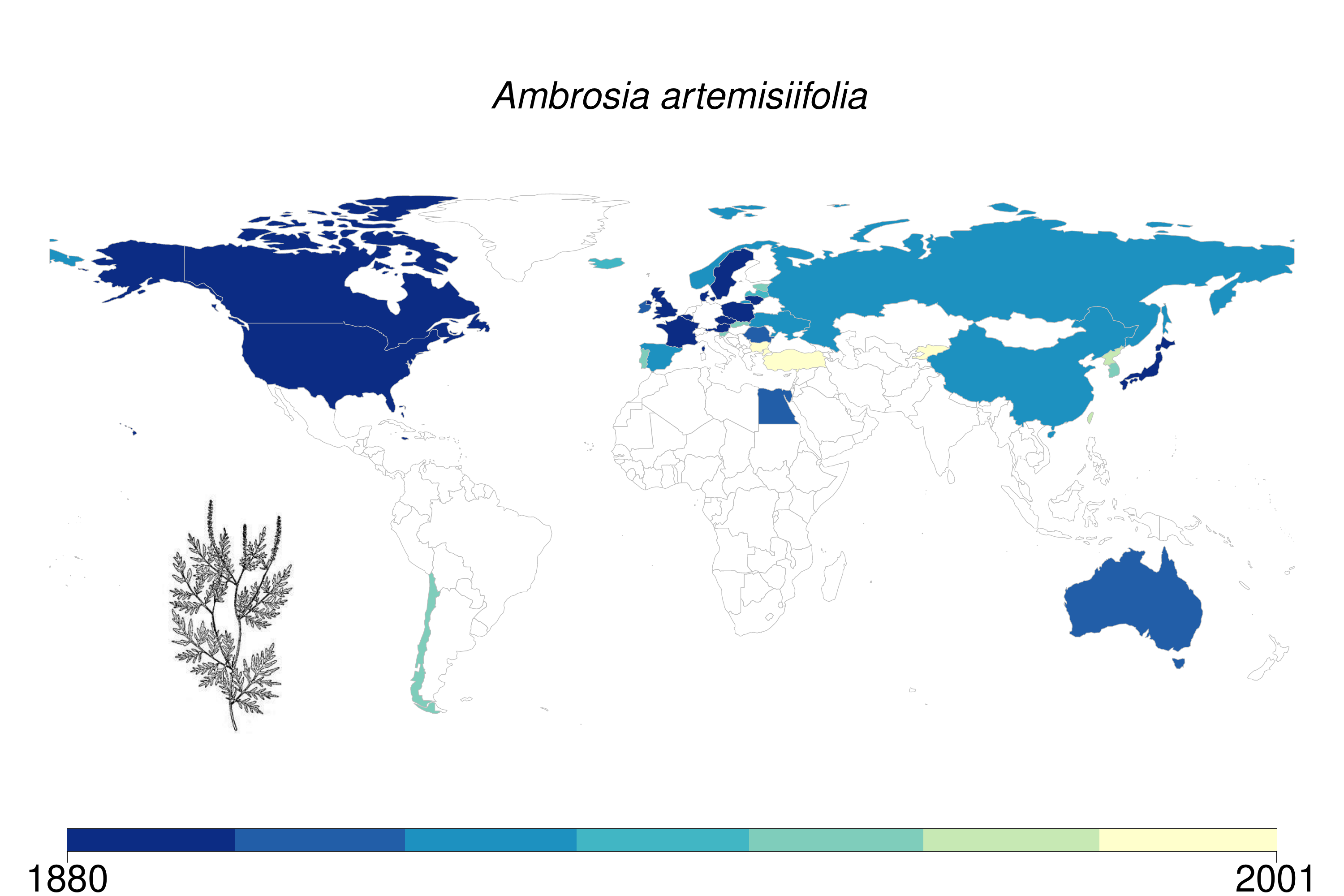}
		\\
			\includegraphics[width=0.5\linewidth]{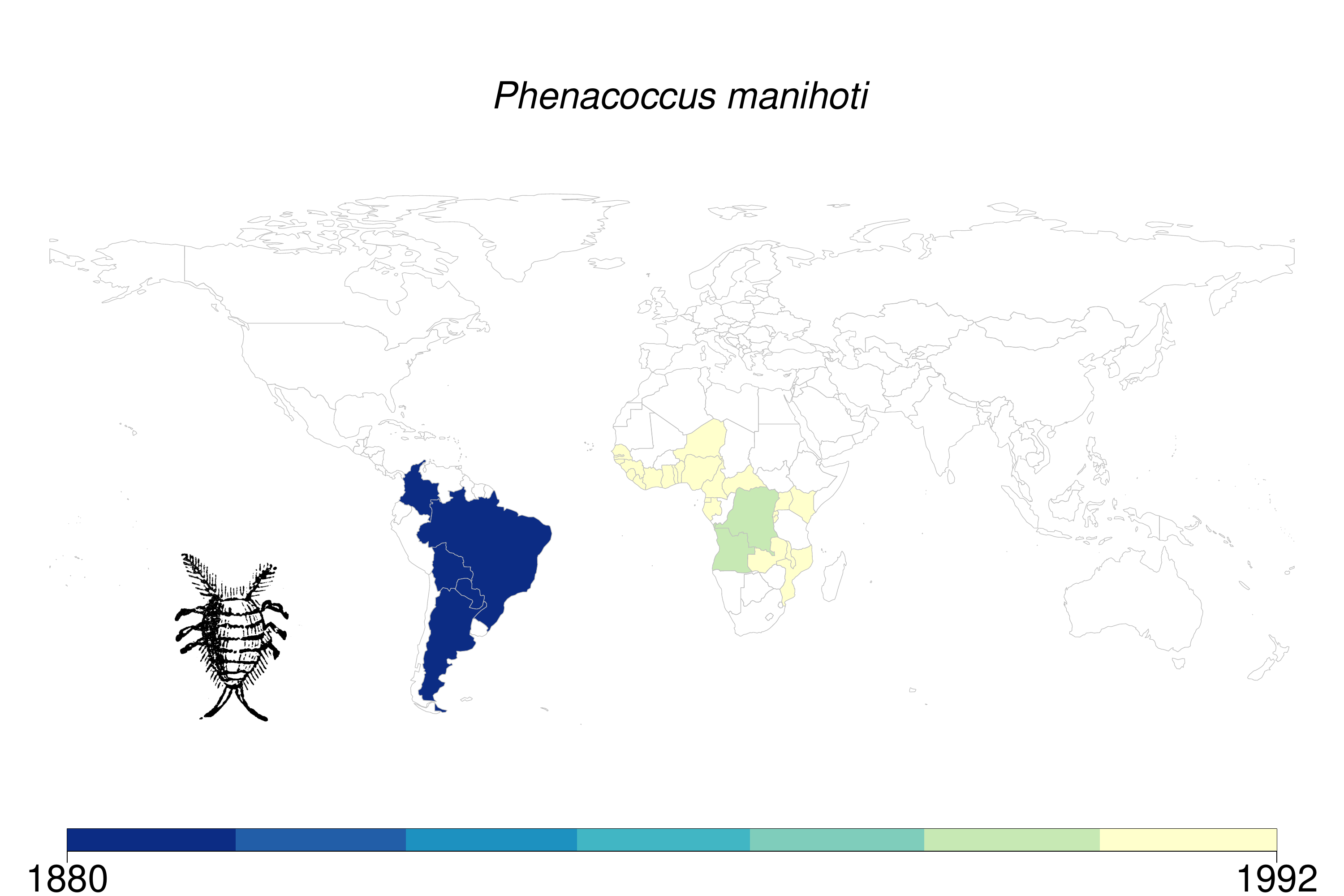}
		&
		\includegraphics[width=0.5\linewidth]{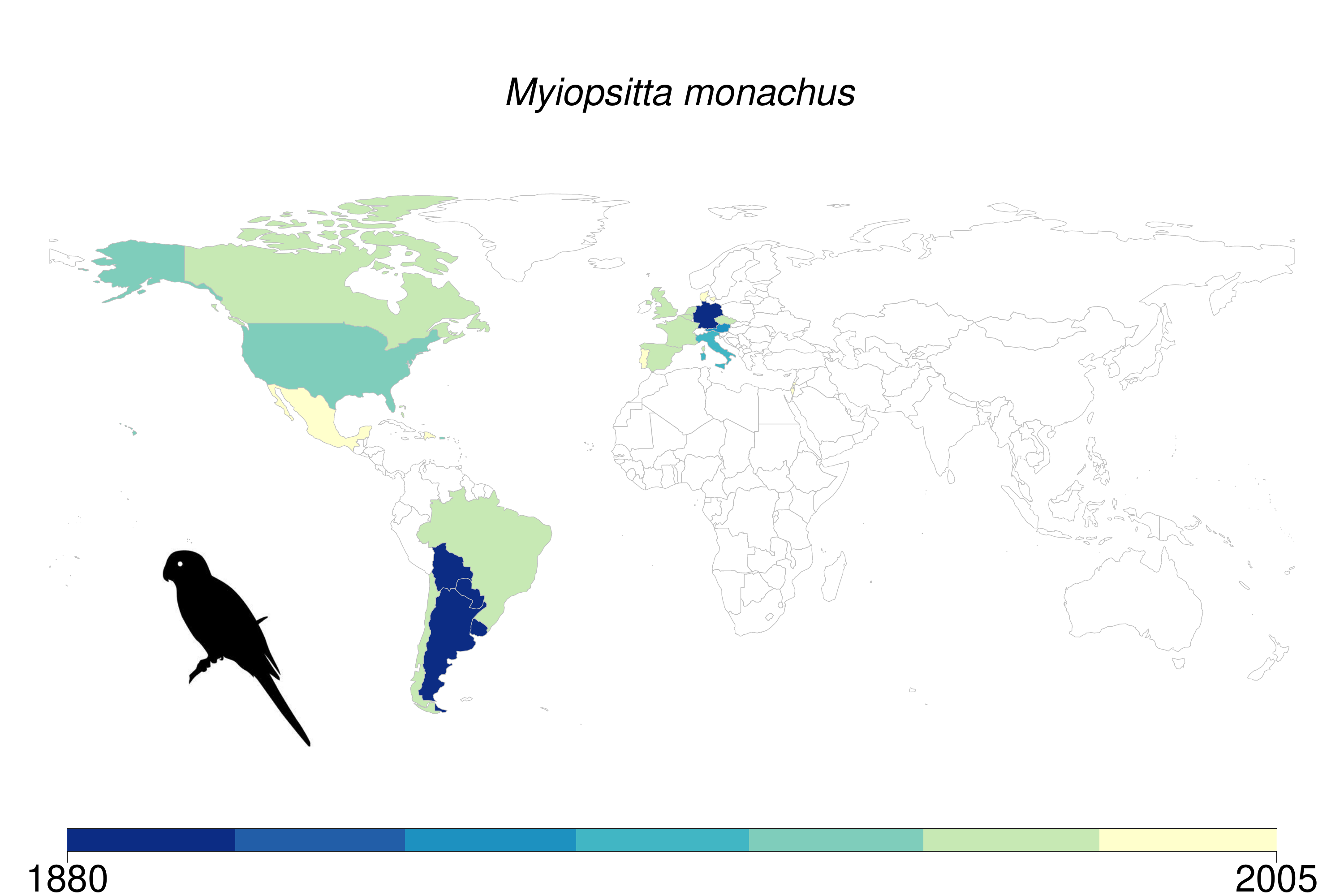}
		\\
	\end{tabular}
	\caption{Spread trajectories of the four species from different taxonomic groups: (top left) muskrat, (top right) common ragweed, (bottom left) cassava mealybug, (bottom right) monk parakeet. Countries are graded from the country invaded earliest (darker) to the latest (lighter). Black colour represents area originally occupied by the species in 1880. Countries without the presence of the species are shown in white.} 
	\label{fig:species}
\end{figure}

For each species, the final data set used for analysis contained the native and alien ranges with the respective first records. The final dataset included 16403 invasion events recorded for 4835 species (615 birds, 186 mammals, 3920 plants, and 114 insect species). The average number of invasion events per taxonomic group was $\sim5$ records per insect species, $\sim4$ per bird and mammal species, and $\sim3$ recorded events per plant species. On average, $\sim46$ invasion events were recorded per region, ranging from 1 to 1685 events (Figure \ref{fig:num_inv}). 157 regions had less than 15 observations, with 33\% of these regions located in Africa; 36 regions had more than 100 records. The region with the highest number of invasion events was Australia. Figure \ref{fig:species} illustrates the spread patterns of four species from different taxonomic groups: the muskrat (mammal), the common ragweed (plant), the cassava mealybug (insect) and the monk parakeet (bird). These plots give important indications regarding the existence of natural environmental barriers affecting the spread and establishment of alien species. For instance, the cassava mealybug is native to South America, but it started spreading throughout sub-Saharan Africa since its inadvertent introduction into the continent in the early 1970s \citep{parsa2012cassava}. The data supporting the findings of this study is publicly available at \url{https://doi.org/10.6084/m9.figshare.20017691.v1}. We also provide a full source code which is available at \url{https://github.com/Juozaitiene/REM-for-species-invasion}.

\subsection{Relational event model for ecological data}
\label{sec:model}

A relational event model (REM) is a probabilistic model that describes the temporal interactions between a (possibly dynamic) set of sender nodes $S(t)$ and a set of receiver nodes $R(t)$ \citep{butts2008a}. Sender and receiver nodes can either represent elements of the same group, so-called \emph{one-mode networks}, e.g., grooming relationships between a group of primates,  or between elements of different groups, so-called \emph{bipartite networks}, e.g., predation relationships between predators and prey. In contrast to traditional network analysis, the relational event model explicitly considers these networks to be embedded in time. There are two main ways in which ecological relationships can be embedded in time. First, the duration of an ecological relationship can be an important and interesting aspect of the relationship itself. For example, in certain animal species, there are sequential periods of monogamy, whose durations may vary. Second, the temporal positioning of the interaction, rather than its duration, can be the primary focus of interest, as for grooming events. The relational event model is intended for modelling interactions of the  second type. 

In this section we introduce the relational event model. First, we focus on a generic model formulation for any type of ecological data. The aim is to explain the idea behind the model and how it can be estimated from the data. Thereafter, we describe a specific implementation of the relational event model designed to model species invasions.

\subsubsection{Generic model formulation}
\label{sec:genericmodel}

\begin{figure}[tb] 
	\setlength{\unitlength}{0.14in}
	\begin{minipage}{0.55\textwidth}
		\begin{picture}(26,33)
			\put(4.45,31){\vector(1,0){6}}
			\put(10.55,31){\vector(1,0){7}}
			\put(7,31.2){$1$}
			\put(14,31.2){$2$}
			\put(8.8,32){Distance}
			
			\put(2,30){\vector(0,-1){30}}
			\put(0.4,2){$t_2~-$}			 
			\put(3,0.5){\framebox(3,3){}}
			\put(3.2,2.6){A}
			\put(3.7,1.8){\PHtunny}
			\put(3.7,0.6){\PHdove}
			\put(9,0.5){\framebox(3,3){}} 	
			\put(9.2,2.6){B}
			\put(9.7,1.8){\PHtunny}
			\put(16,0.5){\framebox(3,3){}}
			\put(16.2,2.6){C}
			\put(16.7,0.6){\PHdove} 	

			\put(3,11){$\mathcal{R}(t_2)= \{(\mbox{\scriptsize \PHtunny},C),~ (\mbox{\scriptsize\PHdove},A)~,(\mbox{\scriptsize\PHdove},B) \}$}
			\put(3,9){$d_{sc}(t_2)=\{~~~~\textcolor{green}{2}~~~~,~~~~\textcolor{blue}{3}~~~~~,~~~~\textcolor{green}{2}~~~~\}$}
			\put(3,7){$\mbox{Prob} ~= \{~~0.42~~,~~~0.16~~~,~~0.42~~ \}$} 
			\put(12,6.8){\framebox(3.3,5.5)}
			\put(3,5) {$L_2(\beta)= \frac{e^{\textcolor{blue}{3}\beta}}{2e^{\textcolor{green}{2}\beta}+e^{\textcolor{blue}{3}\beta}}$}

			\put(0.4,15){$t_1~-$}			 
			\put(3,13.5){\framebox(3,3){}}
			\put(3.2,15.6){A}
			\put(3.7,14.8){\PHtunny}
			\put(9,13.5){\framebox(3,3){}} 	
			\put(9.2,15.6){B}
			\put(9.7,14.8){\PHtunny}
			\put(16,13.5){\framebox(3,3){}}
			\put(16.2,15.6){C}
			\put(16.7,13.6){\PHdove} 	

			\put(3,24){$\mathcal{R}(t_1)= ~\{~(\mbox{\scriptsize \PHtunny},B)~,(\mbox{\scriptsize \PHtunny},C), (\mbox{\scriptsize\PHdove},A),(\mbox{\scriptsize\PHdove},B) \}$}
			\put(3,22){$d_{sc}(t_1)=\{~~~~~\textcolor{blue}{1}~~~~~,~~~~~\textcolor{green}{3}~~~~,~~~~\textcolor{green}{3}~~~~,~~~~\textcolor{red}{2}~~~~\}$}
			\put(3,20){$\mbox{Prob}~ = ~\{~~~0.61~~~,~~0.08~~,~~~0.08~~,~~0.23~~ \}$} 
			\put(8.2,19.8){\framebox(3.8,5.5)}
			\put(3,18) {$L_1(\beta)= \frac{e^{\textcolor{blue}{1}\beta}}{e^{\textcolor{blue}{1}\beta}+2e^{\textcolor{green}{3}\beta}+e^{\textcolor{red}{2}\beta}}$} 
			
			\put(0.4,28){$t_0~-$}			 
			\put(3,26.5){\framebox(3,3){}}
			\put(3.2,28.6){A}
			\put(3.7,27.8){\PHtunny}
			\put(9,26.5){\framebox(3,3){}} 	
			\put(9.2,28.6){B}
			\put(16,26.5){\framebox(3,3){}}
			\put(16.2,28.6){C}
			\put(16.7,26.6){\PHdove} 	
		\end{picture} 
	\end{minipage}%
	\begin{minipage}{0.45\textwidth}
		\includegraphics[width=\textwidth]{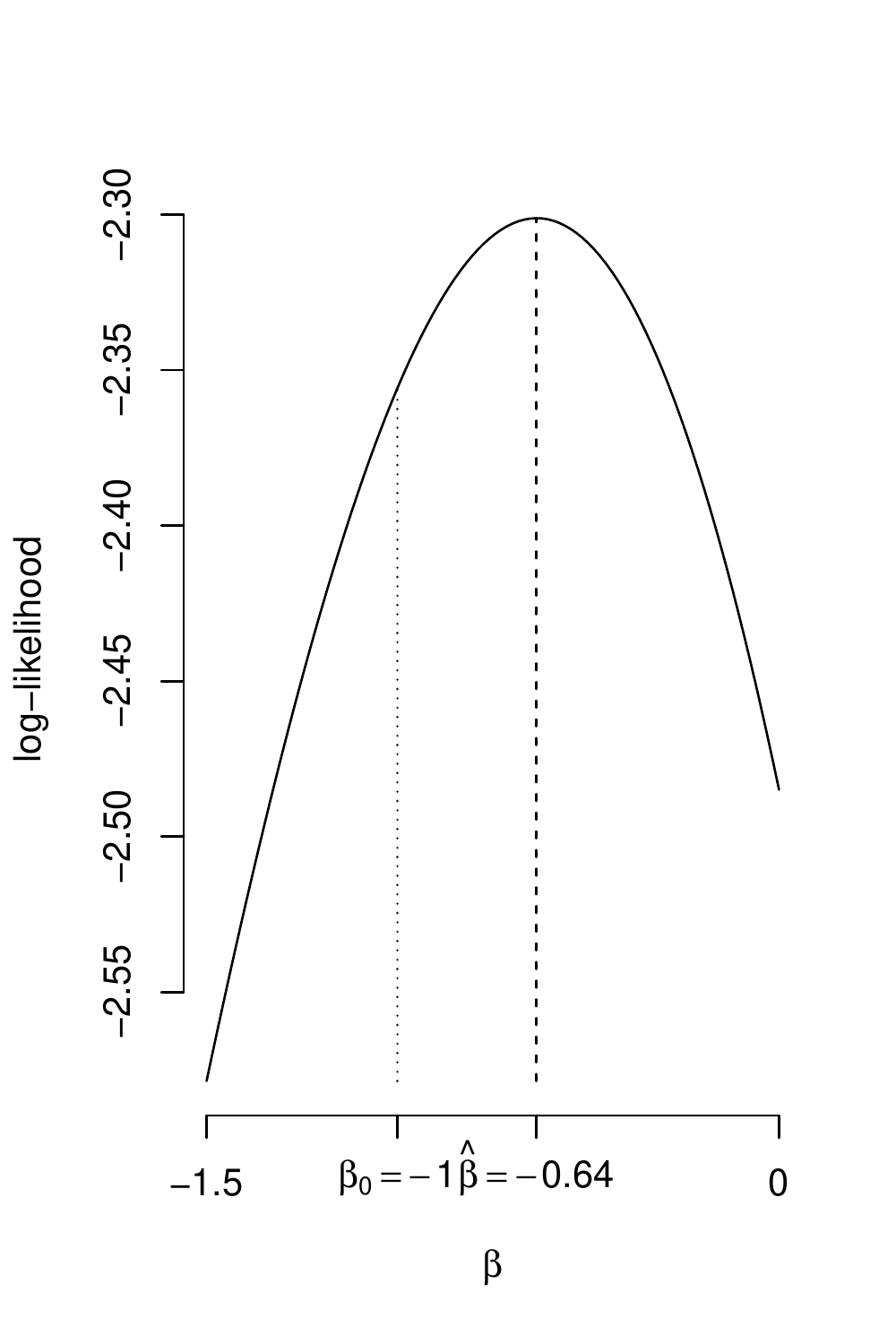}
	\end{minipage}	  
	\caption{An example of an invasion process of two species $\{\mbox{\scriptsize \PHtunny, \PHdove}\}$ in three countries $\{A,B,C\}$ according to the hypothetical hazard function $\lambda_{sc}(t) = \lambda_0(t)e^{-1 \times d_{sc}(t)}$ assuming that the rate of invasion event depends on the distances to the nearest invaded country $d_{sc}(t)$. The effect size of the distance is negative, i.e., $\beta_0 =-1$, meaning that long distance invasions are less likely. At time $t_0$ species $\mbox{\scriptsize \PHtunny}$ is present in country A and species $\mbox{\scriptsize  \PHdove}$ is present in country C. At time $t_1$, species $\mbox{\scriptsize \PHtunny}$ invades country B and at time $t_2$ species $\mbox{\scriptsize  \PHdove}$ invades country A. $\mathcal{R}(t)$ is the risk consisting of all possible invasion events at time $t$ and $d_{sc}(t)$ is the distance to the nearest invaded region at time $t$ of the corresponding invasion event. The probability of each invasion event is calculated following the formula of the event indicator (see section \ref{sec:genericmodel}). In the figure, observed events are shown by a box around the element in the risk set. To estimate the effect of distance (i.e., parameter $\beta$) we use a partial maximum likelihood approach. The partial likelihood of $i$th invasion event is denoted as $L_i(\beta)$. After observing two invasion events $(\mbox{\scriptsize\PHtunny},B)$ and $(\mbox{\scriptsize\PHdove},A)$ the partial log-likelihood, $\log(L_1(\beta)) +\log(L_2(\beta))$, is maximized at $\hat{\beta}= -0.64$, not far from the true effect size of distance $\beta_0=-1$. \label{fig:rem-visual}}
\end{figure}
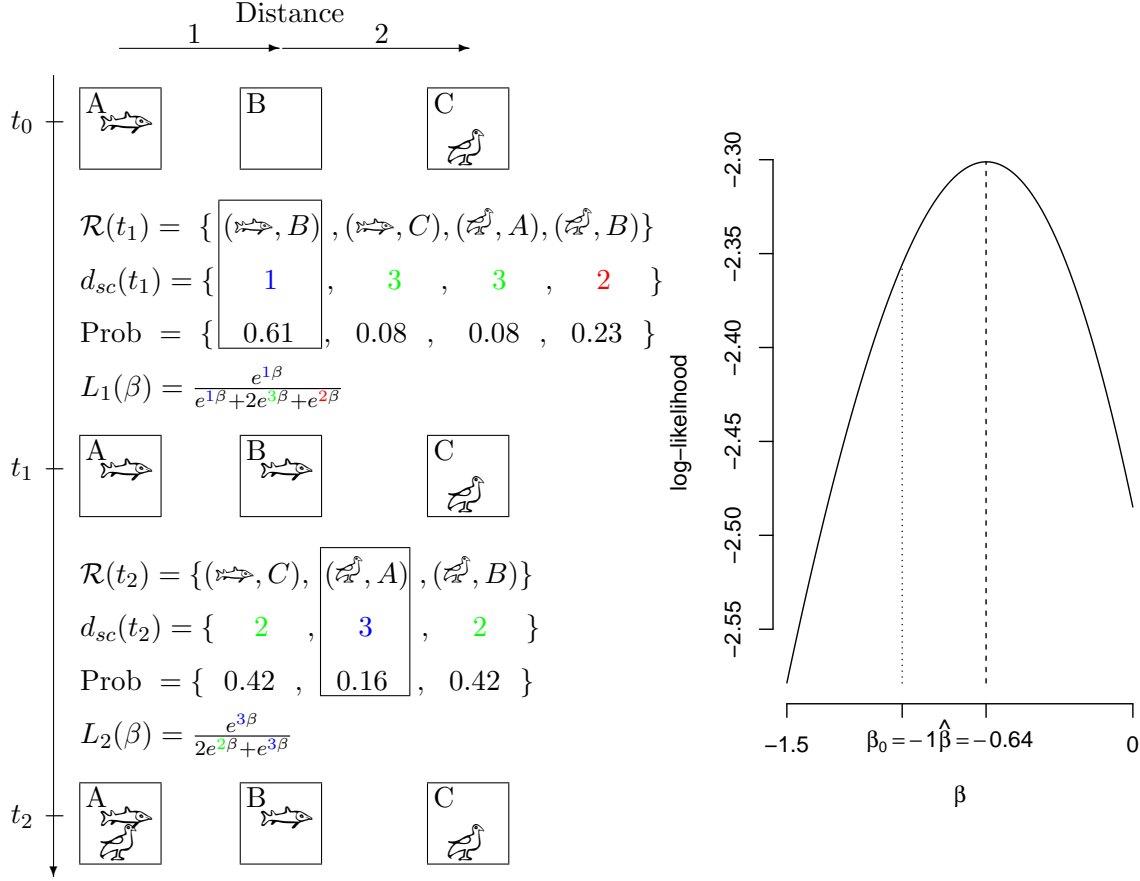

Relational event models are so-called \emph{generative probabilistic models}. This means that they aim to describe the underlying processes that give rise to the observed interaction events between sender and receiver nodes. This is to distinguish these models from \emph{purely phenomenological, statistical models}, which aim to describe the correlations between some observed variables, such as the number of grooming events involving a particular primate in a particular period and its age. In fact, rather than modelling some aggregate quantity, relational event models aim to model each individual event separately. In what follows, we aim to give a description of relational event models, as well as the techniques that are used to estimate them from actual data. An example of all concepts introduced is shown in Figure~\ref{fig:rem-visual}.

A relational event model is considered on a certain time interval $[t_b,t_e]$, where $t_b$ is the beginning and $t_e$ the end of the observation period of the ecological process. During this period interaction events occur between the group of sender nodes and the group of receiver nodes at time instances $t \in [t_e, t_b]$. The set of available senders and receivers can change over time and does not have to be static. This is an important advantage over other network models that typically require the interacting elements to stay the same over the observation period.

\paragraph{Risk set.} As an example, let’s consider a water hole, where groups of prey and predators regularly come to drink. At the beginning of the observation process $S(t_b)$ represents the set of predators and  $R(t_b)$ the group of prey. The predators sometimes attack and kill some prey, while they are gathered to drink. The first attack can be represented as a particular predator $s\in S(t_b)$ that kills a particular prey $r\in R(t_b)$ at a certain time $T_{sr}>t_b$. After that attack, i.e., $t>T_{sr}$, the prey set diminishes, i.e., $R(t)=R(t_b)\backslash\{r\}$, whereas the predator set stays the same, $S(t)=S(t_b)$. However, if a new predator $s_{\mbox{\scriptsize new}}$ joins the group at time $t$, then  $S(t)=S(t_b)\cup \{s_{\mbox{\scriptsize new}}\}$

At any time $t$, the possible interactions $(s,r)$ that can occur is an element of the so-called \emph{risk set} 
\[\mathcal{R}(t) \subset S(t)\times R(t).  \]
For each element $(s,r) \in \mathcal{R}(t)$, there exists, in principle, the event time $T_{sr}>t$, defined as the first interaction time after $t$ between $s$ and $r$. However, only the first of these event times
\[ T = {\arg \min}_{(s,r)\in\mathcal{R}(t)} T_{sr}\]
is the one that is actually observed. After that, as we have seen above, the risk set may change and certain event times may be censored or new ones may be introduced. 

\paragraph{History.} There is another important way that the situation potentially changes after an event occurs. If some predator $s$ has killed prey $r$ at time $T=t$, then we can hypothesize the system to change in two ways: possibly the predator will be satisfied for a while, and the time $T_{sr'}$ that he will attack another prey $r'\neq r$ will be delayed; moreover, the other prey may be spooked and more at their guard, thereby delaying the attack times $T_{s'r'}$ even from other predators $s'\neq s$. To describe the collection of previous events, we introduce the concept of the \emph{history of the process},
\[ \mathcal{H}_t = \{\mbox{all time-stamped interaction events $(s_i,r_i, t_i)$ by time $t$, i.e., $t_i\leq t$} \}\] 
Related to this is the set $\mathcal{H}_{t-}$, which stands for the history of all events strictly before time $t$.

\paragraph{Rate function.} With the above definitions of risk set and history in place, it is now possible to define a generic and generative model for the interaction event sequences. Although this is formally done through the definition of an underlying counting process \citep{perry2013point}, we use a more intuitive description via the sequence of events $\{ (s_i,r_i,t_i)~|~i=1,\ldots,n\}$. The dependence of event times on the constantly changing risk set and history can be expressed by assuming that each event time $T_{sr}$ has a generalized exponential distribution with (non-constant) \emph{rate function} $\lambda_{sr}(t|\mathcal{H}_{t-})$. The rate function should be chosen in such a way as to express the scientific hypotheses about the ecological interaction process that the ecologist would like to test. Given that the rate function is strictly positive, it is common to write the function in an exponential form \citep{cox1972regression},
\begin{equation} \lambda_{sr}(t|\mathcal{H}_{t-}) = \lambda_0(t)e^{x_{sr}(t)'\beta}, \label{eq:hazard} \end{equation}
where $x_{sr}(t) \in \mathbb{R}^p$ are $p$ known \emph{dyadic} covariates at time $t$ that are allowed to depend on $\mathcal{H}_{t-}$, the history of the process right before time $t$. The parameter $\beta\in\mathbb{R}^p$ express the effect size of each of the above covariates. The aim of an empirical ecological study is typically to use the data to test the significance, sign and size of the effect sizes of $\beta$ in order to test one or more ecological hypotheses of interest. The \emph{baseline hazard} is an unknown function of time, which aims to capture residual effects not strictly related to the sender and receivers.

For example, in the predator-prey example, we hypothesized that the rate of the kill process might depend on the appetite of the predator and the alertness of the prey. If we define $x_s(t)$ as the last time before $t$ that predator $s$ killed a prey, then we could hypothesize the following rate function,
\[ \lambda_{sr}(t|\mathcal{H}_{t-}) = \lambda_0(t)e^{\beta_1 \left(t-x_{s}(t)\right) + \beta_2 \left(1-\max_{s'} x_{s'}(t)\right)}, \]
where $\beta_1$ describes the saturation process of the predator and $\beta_2$ the alertness of the prey. Testing whether $\beta_1=0$ or $\beta_2=0$ would allow the scientist to decide whether any of these two ecological hypotheses are real. The baseline hazard $\lambda_0(t)$ may describe a diurnal cycle, capturing the fact that possibly predators are more active in certain times of the day than others. 

\paragraph{Effect estimation from data.} Using the standard properties of the exponential distribution, it can be shown that each inter-arrival time $t_i - t_{i-1}$ has an exponential density with rate $\int_{t_{i-1}}^{t_i} \sum_{(s,r)\in\mathcal{R}(t)}\lambda_{sr}(t|\mathcal{H}_{t-})~dt$, and, more importantly, that the conditional probability of event $(s_i, r_i)$ at time $t_i$ is given by $\frac{\lambda_{s_i r_i}(t_i|\mathcal{H}_{t_i-})}{\sum_{(s,r) \in \mathcal{R}(t_i)}\lambda_{sr}(t_i|\mathcal{H}_{t_i-})}$. This ratio is the result of the relative size of the hazard rate of actual event $(s_i,r_i)$, in the numerator, and the hazards rates from all other possible events $(s,r)$ that could happen at time $t_i$, in the denominator. Using the exponential hazard formulation from (\ref{eq:hazard}), the baseline hazard $\lambda_0(t_i)$ cancels, resulting in
\[ P\left((s_i,r_i) \mbox{ happens }|~ t_i\right) = \frac{e^{x_{s_ir_i}(t_i)'\beta}}{\sum_{(s,r) \in \mathcal{R}(t_i)}e^{x_{sr}(t_i)'\beta}}. \]  
The effect $\beta$ can be estimated by maximizing the probability of all the occurred events, i.e.,
\[ \hat{\beta} = {\arg\max}_\beta \prod_{i=1}^n \frac{e^{x_{s_ir_i}(t_i)'\beta}}{\sum_{(s,r) \in \mathcal{R}(t_i)}e^{x_{sr}(t_i)'\beta}}. \]
This $\hat{\beta}$ is typically referred to as the partial likelihood estimate, which possesses many desirable properties \citep{cox1975partial}.  The baseline hazard can be estimated posthoc according to \cite{Breslow_BioC_75}.

\subsubsection{Relational event model for first invasion events}
\label{sec:invasionmodel}

In this section, we will define the relational event model for first invasion events of alien species into new regions. Here, the sender modes are species and the receiver nodes are regions, and an invasion event describes the first record of an alien species in a region as provided in the FirstRecords database. In principle, it is also possible to consider senders and receivers as geographical units, but as information about the region, where the species came from (i.e., the sender), is lacking, we could not consider this approach. The main aim of applying REM to biological invasions was to disentangle the generative process of the ``first invasion'' relationship between species and regions over time. We analyse four groups of species, $L_i=\{s_1,\ldots, s_{n}\}$ ($i$=mammals, birds, plants, insects), separately across a fixed time-frame $[1880,2005]$, in which the invasion process of the species is recorded relative to geographic set $C=\{c_1,\ldots,c_{272}\}$ of the 272 pre-specified regions. If at time $t$ a species emerged in a new region $c$, then this invasion event can be described by a triple $e=(s,c,t)$, whereby species $s$ spread to a region $c$ at time $t$. The relational event model describes a general framework for modelling and inferring the stochastic species invasion  process $\{(s,c,t)\}_{t\in [1880,2005]}$ in order to identify the reasons why particular species invade particular regions and why they did so at the time they did. Although we study many species of a given taxonomic group together, this does not mean that the relational event model assumes that the species are physically interacting. In fact, the joint mixed effect analysis is able to identify global drivers that are common to the spread of species as well as capture various aspects of individual species dynamics.

We focus on several possible drivers of invasions that can be distinguished into four groups (see Table~\ref{tab:variables}). The first two groups encompass two types of dynamic processes. The first group consists of time-varying factors (i.e. drivers whose values change over time) with constant effects over time. This is expressed by constant coefficients $\beta$ in the hazard function. This group comprises temperature, colonial ties and the number of prior invasions. The second group focuses on, again time-varying, factors whose effects change over time. For example, not only can the volume of international trade vary over time, but the effect of trade volume on the rate of invasion events may also vary over time. This is expressed by time-varying coefficients in the REM. This group encompasses distance, land cover and trade. Whether an effect should belong to the first or the second group depends on whether or not the evidence in the data suggests that a constant effect is sufficient.
The third group covers random effects of a monadic type, which are variables measured at the level of a region or species (species invasiveness and region invasibility), whereas the final group of factors considers dyadic random effects, which denote variables measured at the level of pairs of species, capturing the inter-dependence in alien species spread. 

For each taxonomic group separately (plants, birds, insects, mammals), we fit a relational event model representing global invasion dynamics over the period from 1880 to 2005. The overall time-dependent hazard function of species invasion dynamics considered for each taxonomic group is given as,
\begin{eqnarray}
	\log \lambda_{sc}(t) &=& \log \lambda_0(t) + d_{sc}(t)\beta_1(t) + tr_{sc}(t)\beta_2(t) + dt_{sc}(t)\beta_3 +  l_c(t)\beta_4(t) \nonumber  \\ &&  + u_c(t)\beta_5 + k_{sc}(t)\beta_6 + pi_{c}(t)\beta_7   + se_{c}\beta_8 + b_s +b_c + b_{ss_c(t)},  \label{eq:model}
\end{eqnarray}
where 
\begin{itemize}
	\item The baseline hazard $\lambda_0(t)$ captures the underlying changing intensity of the overall invasion process.
	\item $d_{sc}(t)$ denotes the distance to the nearest region $c$ in which species $s$ is present by time $t$. The proposed definition accounts for the distances between invaded countries as well as native ranges. Long-distance invasion events through natural dispersal are typically rare. Therefore, the distance between two regions that species have to overcome could be an important factor responsible for alien species invasions. The distance between two regions is defined as the distance between their closest borders, and thus the distance between neighbouring regions is zero. The proposed distance covariate is time-varying, as the minimum distance between a region and another invaded region will change over time as new regions get invaded. An example is shown in Figure~\ref{fig:rem-visual}, which also explains the estimation procedure.
	\item The time-varying quantity $tr_{sc}(t)$ is the logarithm of the sum of annual trade flows between region $c$ and other regions invaded by species $s$ by time $t$ -- the log-transformation is used to capture saturation. 
	\item The variable $dt_{sc}(t)$ represents the minimum absolute difference in temperature between region $c$ and other regions invaded by species $s$ by time $t$. It is a time-dependent variable since the list of regions invaded by species $s$ is changing over time.
	\item $l_c(t)$ is the time-varying variable that describes the proportion of agricultural land, defined as the sum of cropland and pasture proportions in the region $c$ at time $t$. 
	\item $u_c(t)$ is the proportion of urban area in region $c$ at time $t$. 
	\item $k_{sc}(t)$ indicates the presence of species $s$ at time $t$ in a region within the area of a colonial power. 
	\item The variable $pi_{c}(t)$ is a weighted number of prior invasions in region $c$. We use a standard inverse exponential downweighing of recent invasions to capture the current invasibility of certain regions for alien species invasions, i.e, $pi_c(t) = \sum_{s\in L_i } 1_{\{t_{sc}<t\}} {0.95}^{t-t_{sc}}$, where $t_{sc}$ is the moment that species $s$ invaded region $c$. This variable summarizes current invasibility by giving more weight to more recent events.
	\item The variable $se_c$ aims to correct for confounding effects by capturing the differential sampling effort within each country. It is defined as the number of species that were recorded in each country by 1880.
	\item The relational event model also includes random effects representing species invasiveness ($b_s$) and region invasibility ($b_c$). It is expected that different species may vary in their spread behaviour, whereas similarly
certain regions may be more attractive destinations than others.
	\item The last group of model explanatory variables considers dyadic random effects that aim to capture interdependence in alien species spread.  The effect $b_{ss'}$ describes the affinity of species $s$ and $s'$ in their spread patterns. The species indicator $s_{c}(t)$ is defined as the last species to invade $c$ before $t$.
We consider both a general formulation in which the order of invasion by the species is accounted for, as well as the more parsimonious symmetric formulation that constrains $b_{ss'}=b_{s's}$. It turns out that although the symmetric formulation is often sufficient, for some taxonomic groups the general non-symmetric random effect offers additional explanatory power. However, some species combinations are very rare or never appear in the data. To robustly estimate the effect of species interdependence in spread, which is not affected by low invasion counts, we focus on species with the highest number of recorded invasion events. Thus, this specific effect is estimated using only the information on the top thirty most widespread alien species in each taxonomic group.
\end{itemize}

Time-varying effects were modelled by estimating a piecewise constant coefficient for each of five equally sized periods  (1880-1905, 1906-1930, 1931-1955, 1956-1980, and 1981-2005). Analyses were conducted using the \textit{coxme()} function from the {\tt coxme} package version 2.2-16 within the R statistical software  to fit a Cox proportional hazard model with mixed effects \citep{R, coxme}. Chi–square tests were used to compare different models. To assess whether a fitted Cox regression models adequately described the data we performed model diagnostics. The proportional hazards assumption was checked using statistical tests and graphical diagnostics based on the scaled Schoenfeld residuals. Thus, we used the \textit{cox.zph()} function from the {\tt survival} package version 3.2-13 \citep{survival-package} to test for independence between residuals and time, the \textit{ggcoxzph()} function from the {\tt survminer} package version 0.4.9 \citep{survminer} to perform a graphical diagnostics and the \textit{ggcoxdiagnostics()} function from the {\tt survminer} package version 0.4.9 to test influential observations or outliers. The diagnostic tests suggested an overall adequate fit of the model, for more details see Supplementary Materials (Section S2). A more detailed description of the inference and model diagnostics is provided in the online supplementary materials.

\section{Results}
\label{sec:results}

In this section, we present the results of the alien species invasion analysis using the relational event model. 
The final REMs provide a parsimonious description of the data as summarized in Table~\ref{tab:summary}. The models explained between 12.39 and 26.1\% of the variance, depending on the taxonomic groups as shown in Table~\ref{tab:summary} (all p-values $< 10^{-4}$). Note that the provided $R^2$ values should not be confused with the $R^2$ for linear regression. In event history models, such as the REM, the $R^2$ coefficient has a fundamental upper limit of roughly 56\%, if there is no additional over-dispersion and all the fundamental causes and their effect sizes were known perfectly \citep{zheng2000summarizing}. In this light, the reported $R^2$ values suggest that the models provide significant insight in the underlying generative process. To confirm the hypothesis that the magnitude of the effect of some factors may change over time, we tested for the benefit of including time-varying effects. Thus, we analysed an extensive list of different models, where the smallest model included only static coefficients and the largest included all time-varying coefficients. To select the best model for each taxonomic group we performed Wald tests for simple effects and chi-squared tests for time-varying effects. Goodness-of-fit tests presented in Table \ref{tab:comparison} show that the proposed models with several time-varying coefficients are a significant improvement over the model including only static coefficients as well as the model including only time-varying coefficients. More detailed results are provided in the Supplementary Materials (Tables S1-S2). The performed analysis reveals that models that include distance, land cover and trade are highly significant, with the sole exception of cropland-pasture effects for insects.    

\begin{figure}[h]
	\centering
	\includegraphics[width=0.7\linewidth]{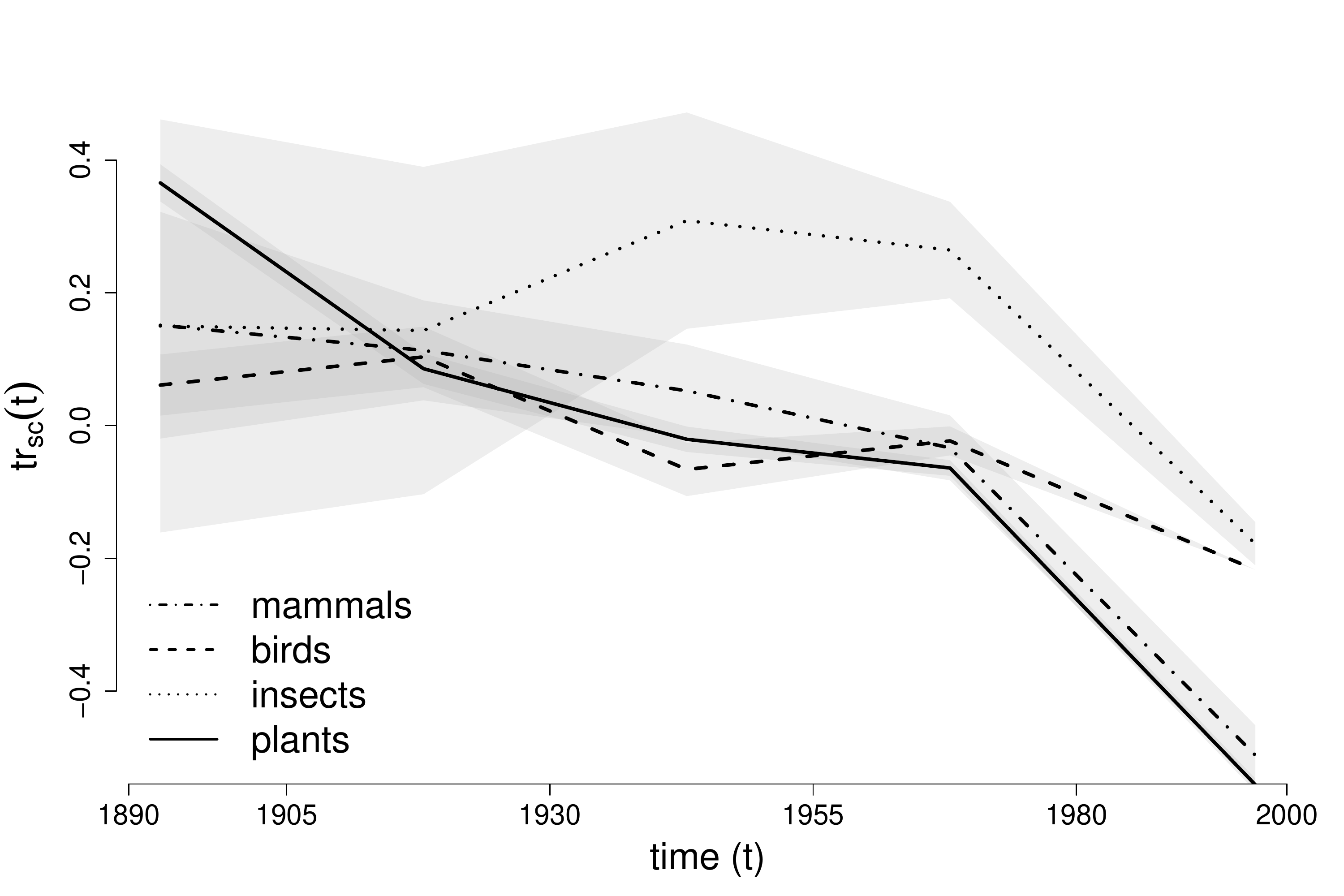}
	\caption{Effect of trade on the process of alien species spread. After an initial positive effect, in more recent times the effect of trade seems to have become negative. The grey shade area represents the 68\% confidence interval. For all taxonomic groups the null-hypotheses of no time-varying trade effect can be rejected, as all the associated p-values $<0.001$.}
	\label{fig:trade}
\end{figure}

\paragraph{Trade flow.} It is well-established that trade is crucial for alien species spread across the globe \citep{seebens2015global,dawson2017global,essl2011socioeconomic}. However, this picture seems to be more complicated for invasion events than might be expected. Although trade was certainly of major importance in the early 20th century, the effect seems to have largely disappeared more recently. During the first two periods (1880-1930), the effect of trade was positive, i.e., the spread rate increased with the amount of trade between a region (Figure~\ref{fig:trade}). However, the effect has become negative since. This finding may seem counter-intuitive, but can partially be explained by the fact that the number of introduced species did not keep pace with the increase in trade volumes due to a saturating relationship between imported commodities and the number of alien species \citep{Seebens2017}. As a consequence, the relative importance of trade as a predictor for alien species invasions declined over time. Further, the proportion of goods with known high relevance for introducing alien species such as agricultural and silvicultural goods (e.g. wood, grain) has declined severely in the last decades \citep{luppold1988hardwood}. There has also been a structural shift in commodity composition, e.g., bulk commodities have been substituted by processed consumer-ready products. The share of bulk commodities declined from 50\% in 1980 to 32\% in 1995, while the share of non-bulk products increased from 50\% to 68\%. Moreover, the share of products with low risks of introducing alien species (e.g. electronics, goods from other technology-intensive industries, and labour-intensive products, particularly clothing) have grown most rapidly in world exports during the period 1980-2000 \citep{mayer2003dynamic}. A broader analysis of the absolute values of traded goods would give a deeper understanding of the development of international trade flows. However, it is very difficult to establish to what extent the absolute trade (in units of weight such as tons) of raw goods has changed over the past century. Even the value of bulk trade is not a good proxy and, even then, is only known for a few instances. 

\begin{figure}[h]
	\centering
	\begin{tabular}{cc}
		\includegraphics[width=0.5\linewidth]{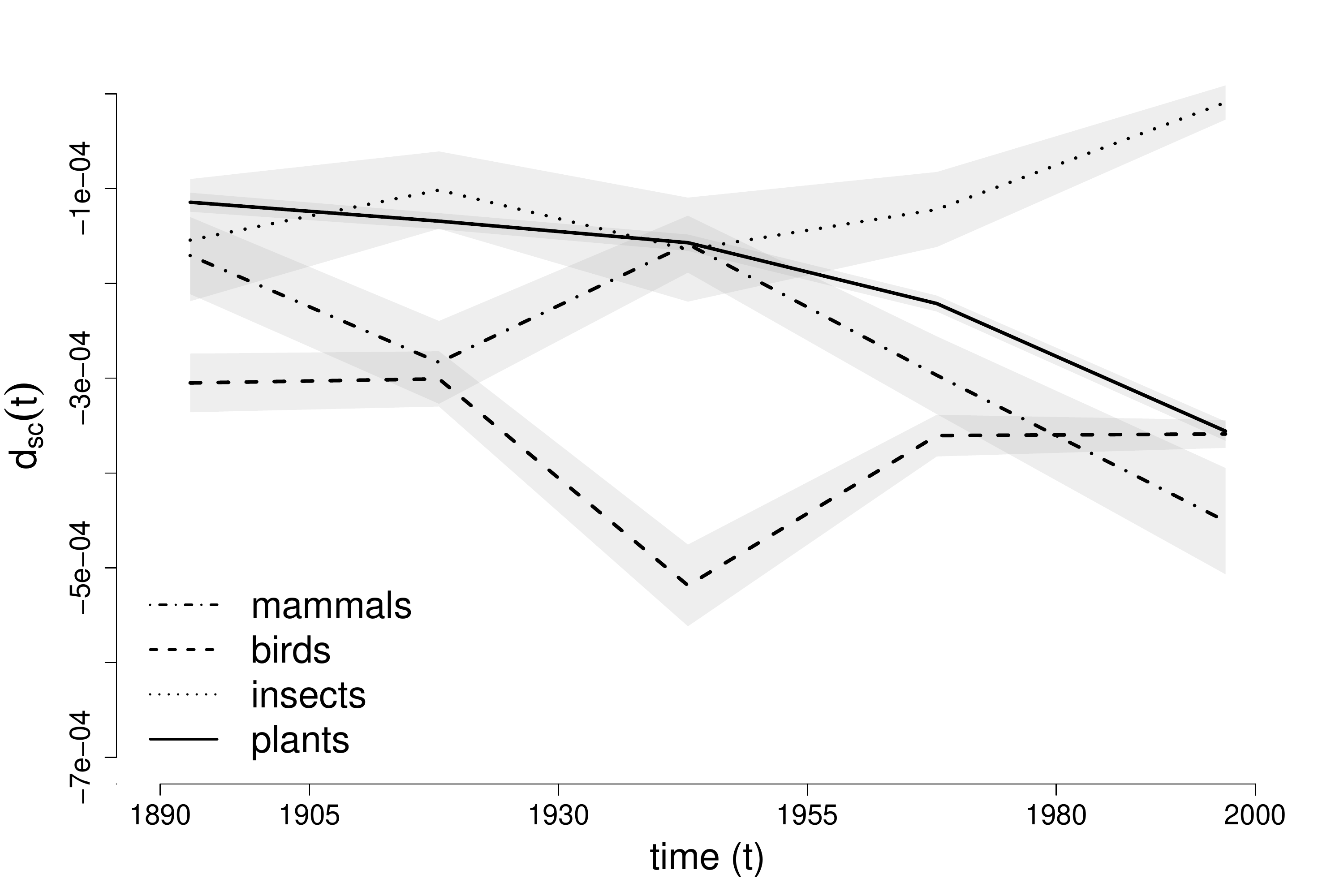}
		&
		\includegraphics[width=0.5\linewidth]{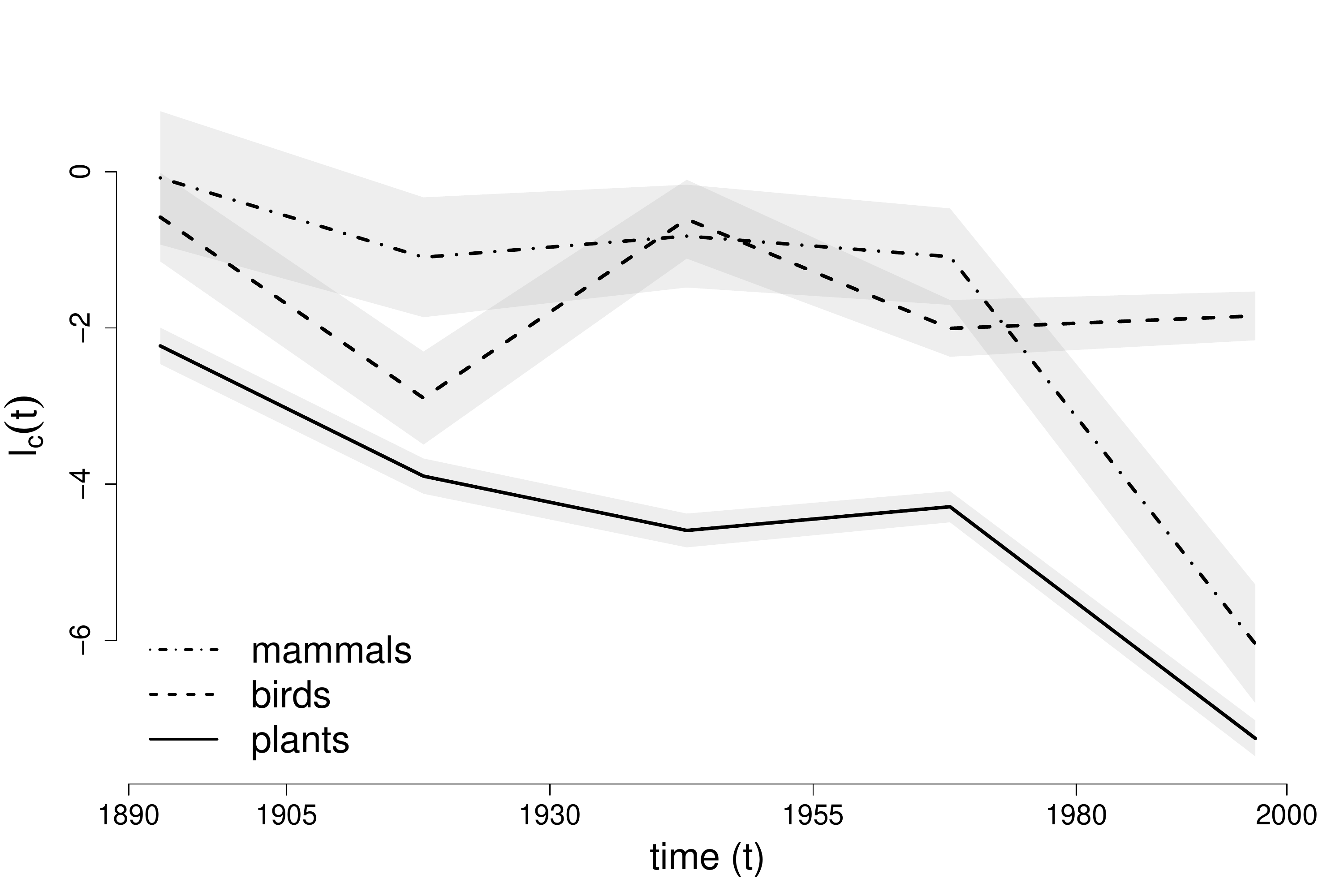}
		\\
		(a) & (b)
	\end{tabular}
	\caption{ Effect of a) distance and b) cropland and pasture proportion on rates of invasion. a). Negative effect of distance on the rate of invasion events has increased over time for mammals and plants. For all taxonomic groups p-value $<0.001$. b). Crop land and pasture proportions have negative effects on the rate of invasion events. The grey shade area represents the 68\% confidence interval. For all taxonomic groups the null-hypotheses of no time-varying distance and land cover effects can be rejected, as all the associated p-values $<0.001$.} 
	\label{fig:distance}
\end{figure}

\paragraph{Distance.} Distance has a negative effect on the rate of invasion events, which means that long-distance invasion events are typically rare. Moreover, for most taxonomic groups the effect size grew over time, i.e., the importance of distance on species invasions increased in recent years (Figure~\ref{fig:distance}a). We argue that this can possibly be explained by the development of national and international legal instruments focusing on reducing the risk of introducing alien species through transportation \citep{turbelin2017mapping}. The effect of distance has become especially strong after 1980 for mammals and plants, which coincides with the fact that since the 1990s there has been a significant increase in legislation regarding invasive alien species, especially in countries with high rates of species invasion. The opposite trend has been inferred for insects, i.e., the effect of distance decreased over time but has remained negative.

\paragraph{Land cover variables.} The effect of the proportion of urban area in region is only significant for vascular plants. This effect has been consistent over the whole time span we analysed (Table \ref{tab:est}). Increases in the fraction of urban areas in a region are negatively correlated with spread rate for plants, as shown by the negative coefficients. More specifically, for every increase in urbanization by 1\%, the rate of invasion events in that region decreases by 16.8\%. On the other hand, the effect of cropland and pasture on the rate of invasion events is more complex and has changed over time in a non-linear fashion. It is significant for plants, mammals and birds and its time-varying coefficients is negative and has been growing for most of the period under consideration. Recently,  this effect for birds seems to be decreasing again (Figure~\ref{fig:distance}b). This means that larger fractions of cropland and pasture in a region resulted in lower spread rates. Modern agricultural techniques may be responsible for these particular changes.

\paragraph{Sampling effort.} The proxy for sampling effort has a significant effect on the rate of species invasions for plants, insects and mammals, confirming previous studies by \cite{bonnamour2021insect}. For our purposes, sampling effort has shown to be an important confounding effect that should be accounted for in the model, in order to correctly interpret other structural effects. 

\paragraph{The number of alien species found in a particular region.}
Both plants and insects are significantly affected by the number of species that have already invaded a certain region. Both plants and insects were recorded less frequently in regions that had already been invaded by many other species. The effect for plants was nonetheless very small but statistically significant, i.e., decreasing the hazard by 0.5\% per invasion. For insects the observed effect was higher previous invasions decreased the hazard by 10.4\%.

\paragraph{Thermal similarity.} For all taxonomic groups, rates of invasion events are significantly affected by the climatic conditions of the receiving region. The temperature differential effect for all taxonomic groups is negative (Table~\ref{tab:est}). The parameter values associated with $dt_{sc}(t)$ mean that for each degree of mean annual temperature difference to the nearest invaded region, the spread rate has decreased by 15.8\%,  8.5\%, 6.2\% and 5.1\% for plants, insects, mammals and birds, respectively, keeping everything else constant. The negative effect of temperature for all taxonomic groups shows that biological invasions are strongly affected by climatic conditions. This finding supports the importance of climate matching between regions for alien species spread and establishment \citep{abellan2017climate,lovell2021environmental}.

\paragraph{Colonial ties.} The effect of colonial ties is not statistically significant for any taxonomic group.

\paragraph{Species invasiveness.}

Analysing the random effects, we can identify the species with the highest invasiveness propensity (i.e. rate of spread across regions after accounting for the influences of covariates). After accounting for the other effects described above, the random invasiveness rate captures the tendency of different species to spread. Table S3 in Supplementary Material presents the top five alien species in each taxonomic group. Siberian Chipmunk (\textit{Tamias sibiricus}), Raccoon dogs (\textit{Nyctereutes procyonoides}), and American mink (\textit{Neovison vison}) were the mammals with the highest invasiveness rate. These results are in line with previous studies \citep{tedeschi2021introduction,Tedeschi2021.04.21.440832} stating that these species are the most widespread in Europe, having invaded at least 27 countries each. The prevalence of these species may be related to the fact that many game species and fur-bearing animals have been moved, introduced, or have even escaped from farms into the wild.

For plants, water plants (\textit{Hydrocotyle ranunculoides}) and conifers such as Caribbean pine (\textit{Pinus caribaea}) and Douglas fir (\textit{Pseudotsuga menziesii}) ranked first. It is argued that conifer species escaping from plantations and becoming invasive are often those that have been cultivated the most widely and for the longest time \citep{kvrivanek2006planting}, while small seed mass, short juvenile period, and short intervals between large seed crops have been shown to contribute to the invasiveness of conifer trees \citep{richardson2004conifers}.

The top three birds species with the highest invasiveness rate were the Rock pigeon (\textit{Columba livia}), followed by the Alexandrine parakeet (\textit{Psittacula eupatria}) and the Egyptian goose (\textit{Alopochen aegyptiaca}). Indeed, Rock pigeons have an extensive distribution in Eurasia and North Africa \citep{goodwin1960comparative}, whereas Alexandrine parakeets and Egyptian geese are considered among the most widespread alien species in Europe \citep{Gyimesi2012,csahin2019breeding}.

The highest ranking insects were pests,  i.e. a fruit fly (\textit{Bactrocera invadens}) which was first introduced in Kenya and has rapidly expanded in 28 African countries \citep{khamis2012taxonomic}; pink hibiscus mealybug (\textit{Maconellicoccus hirsutus}), which is considered a highly invasive species that has been introduced to 75 countries in all over the world \citep{milonas2017pink}, and western flower thrips (\textit{Frankliniella occidentalis}), which is indicated as a major worldwide crop pest \citep{kirk2003spread}.

\begin{figure}[h]
	\centering
	\includegraphics[width=0.7\linewidth]{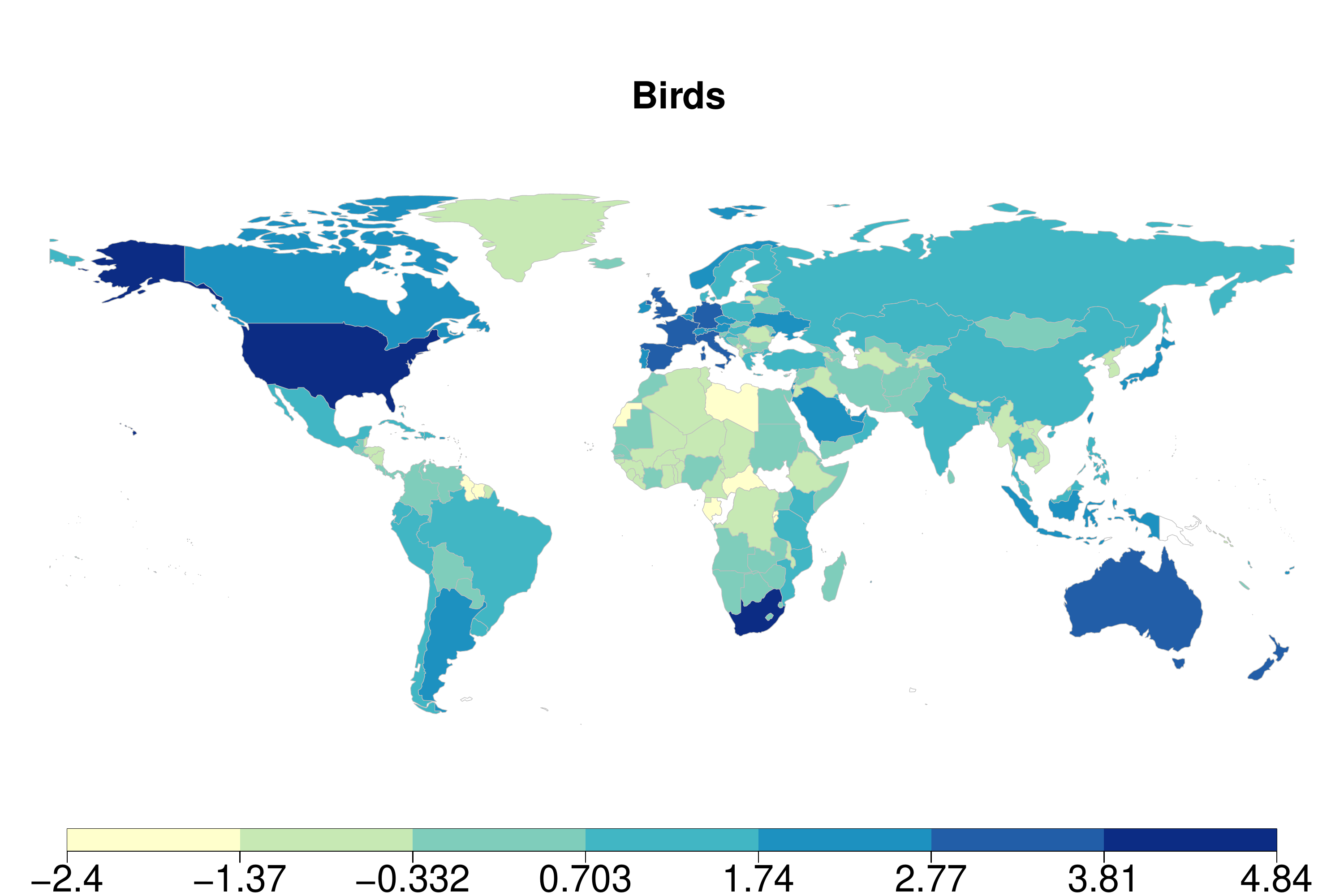}
	\caption{Global map of region invasibility, indicating regions susceptibility to alien species invasions. Countries are graded from the regions with the highest invasion propensity (black) to the smallest (light grey), while correcting for all fixed effects in the REM. United States, South Africa, Australia, New Zealand are geographic hot spots of invasion for birds.}
	\label{fig:cntr}
\end{figure}

\paragraph{Region invasibility.} 
Figure \ref{fig:cntr} represents the random region effects reflecting region invasibility for birds. Plots for other taxonomic groups are provided in the supplementary materials, Fig. S15. Several countries that were part of the former British Empire (New Zealand, South Africa, United Kingdom, USA) are on top of the list of the country invasion propensity list for all four taxonomic groups. This finding is in line with previous findings on hot spots of global invasion levels across eight taxonomic groups, where these countries all show up prominently \citep{dawson2017global}. There are likely several reasons that jointly explain this result. Interestingly, most of these countries are large and have a long history of an advanced ecological-scientific infrastructure that has kept accurate records of biological data including species invasions. Further, these countries have been part of one of the largest European Empires and the integration into a near-global political entity and higher levels of trade and people movements among these regions may have fostered biological invasions as well \citep{lenzner2022naturalized}. On the other hand, several of these countries such as New Zealand and Australia have implemented strict biosecurity policies to stop the introduction of alien species \citep{turbelin2017mapping}. However, these policies are mostly of rather recent times and may be masked by the time frame (1880-2005) under observation here.

\paragraph{Interdependence of alien species spread.}

Standard deviations of the random effects show that the symmetric formulation of the interdependence in spread propensity of pairs of species provides similar explanatory power as the non-symmetric formulation in most cases (Table~\ref{tab:est}). Only for mammals did the non-symmetric random effect offer additional explanatory power. In this case, the temporally symmetric effect for species interactions between mammals is not significant, whereas the temporally asymmetric effect for mammals is highly significant. This clearly shows that temporal ordering is an important aspect in explaining species invasions. Models that would not distinguish between temporal order would not be able to identify this effect. The estimated random effects for insects are represented in Figure~\ref{fig:interaction}, whereas the others are included in the Supplementary Materials. 

\begin{figure}[h]
	\centering
	\includegraphics[width=0.9\linewidth]{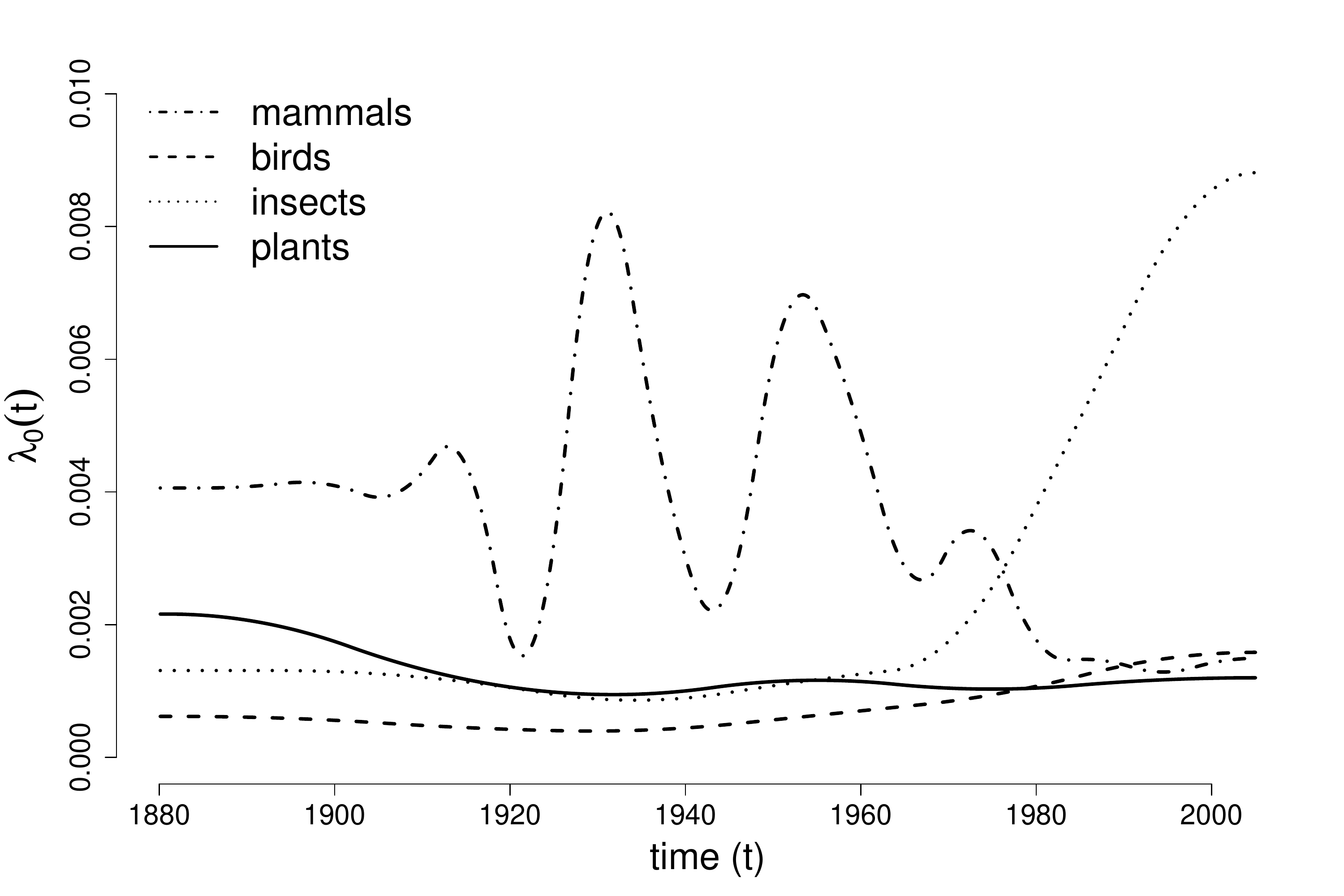}
	\caption{Baseline hazard for the underlying invasion speeds of the different taxonomic groups. Although mammals have shown the quickest pace, it is tapering off recently, whereas the spread of  insects is speeding up since 1980.}
	\label{fig:blh}
\end{figure}

\paragraph{Baseline hazard.} The baseline hazard $\lambda_0(t)$ is a remainder term of the relational event model. It represents the underlying rate of spread over time that is unexplained by the drivers included in the model. Figure~\ref{fig:blh} shows the cumulative baseline hazard estimates for the four taxonomic groups covered. The linear shape of the baseline hazard for both plants and birds suggests that they have an approximately constant baseline hazard $\lambda_0(t) = \frac{d\Lambda_0(t)}{dt}$, indicating that the spread rate as explained by the model was mostly constant for these species. For mammals, the cumulative baseline hazard has flattened since the 1960s, suggesting that their spread rate has been decreasing. Earlier, the spread rate of mammals outpaced that of the other taxonomic groups by a factor of two to three. Interestingly, the two big waves of the mammal curve correspond to the two World Wars, suggesting a real variation in species invasions as a result of the World War (less international trade and travel, except military vehicles), and it is related to a recording artifact since the number of available first records likely decreased because of lower sampling. Instead, we observe a rather constant hazard for birds and plants, suggesting no general increase in the spread rate. However, the picture for insects is quite different, i.e., insects have seen a marked increase in their baseline hazard since the 1980s. This implies that the spread rate has been accelerating for insects since then in a way that cannot be explained by the current variables in the model, which is in line with previous findings by \citep{bonnamour2021insect}. These results partly suggest that international biosecurity policies and legislation have varying levels of success for different taxonomic groups. We hypothesize that preventing and controlling early-stage mammal and bird invasions is much easier than stopping insect invasions. Most species of mammals and birds have been intentionally introduced for sport hunting, commercialization and domestication as livestock, pets, or for pest control \citep{carpio2020intentional}. Some other species have been semi-intentionally introduced (intentional import but accidental escape), while most insect invasions are accidental. This is because most insects are small and often go unnoticed until they have become established \citep{eschen2015taxonomic}.




\section{Discussion}
\label{sec:discussion}

In this section, we will first discuss the complexities and limitations of the application of the relational event model to the alien species invasion process. Furthermore, we discuss the opportunities as well as the difficulties of applying this model more broadly in ecology.   

Relational event models offer the capacity to account for the dynamic nature of ecological relational events, such as biological invasions, and to investigate how relationships between these events and potential drivers change over time. These models have in principle the ability to identify causal drivers of the relational process by accounting for confounding effects that almost certainly are present in any complex ecological scenario. However, REMs require comprehensive and accurate temporal data on potential drivers and events, which are often scarce or difficult to come by. Our analyses should be seen against the backdrop of possible recording delays of the first introduction of a species in a region. Other sources of bias include uneven sampling over space and time as well as species-specific detectability issues. The detectability may be influenced by various factors, such as species richness, body size, and environmental factors, such as vegetation structure or successional change \citep{isaac2011distance}. The uneven sampling of alien species across regions mirrors the uneven distribution of human population and socio-economic activity. Some regions have a long history of an advanced ecological-scientific infrastructure where  recording of biological data including species invasions is much better than in other regions. These uncertainties might explain the comparatively low degree of explained variation. Information on trade, land use, and economic variables was not easily available, and therefore proxies had to be employed. 

Moreover, alien species spread is a much more complex process than the one described in the presented REM, which is based on several simplifying assumptions. For instance, distance can affect post-introduction natural dispersal and also human-mediated dispersal \citep{wilson2009something}. In the case of natural dispersal, the effect of distance on alien species spread is species-specific (because of dispersal abilities) and, in the case of human-mediated dispersal, the effect of distance on spread depends on pathways (e.g., intentional or accidental), vectors (e.g., car or boat, or plane) and also species traits (e.g., attaching capacities, egg laying behaviour, preferred substrate) \citep{meurisse2019common}. Some species might be poor dispersers naturally but easily transported through human-mediated dispersal and other species can spread widely by their own following the initial introduction (e.g., \textit{Vespa velutina}) \citep{verdasca2021}. This means that although our relational event models give a more subtle picture of the time-dependent invasion process of species, the parameter estimates have to be interpreted in the light of these limitations. 

This paper introduced a flexible and powerful REM modelling approach for the analysis of ecological data and demonstrated its practicality and effectiveness by applying the approach to the process of alien species invasion. These models are highly generic and merely require information about the temporal sequence of interactions. Analysis software is available in any statistical package, which means that there are very few practical obstacles for using these models more widely in ecology.

The REM can be generalized to any ecological interaction process where a number of actors interact together over time. Biological interactions, which we more generally refer to as 'relational events', are the fundamental process shaping the diversity of life \citep{Del-Claro}. These relational events capture important basic relationships in ecology \citep{del2021evolutionary}. They can be intraspecific, i.e., such as male-male interference competition \citep{dijkstra2018does} or mating, interspecific, such as predation or pollination, or form some more generalized interaction, e.g., between a species and its habitat or the environment. Often it is possible to associate particular temporal events with these interactions, where both the event type and the timing of the event are relevant. For example, the relational event model is well suited for data structures common in animal behaviour studies: behavioural data collected via proximity loggers provide reliable, highly granular information about social interactions between animals. The REM has been used to model animal social networks in cattle \citep{patison2015time,tranmer2015using}. This model can also be applied to investigate scenarios where social connectivity is related to population outcomes that are linked to social factors such as cooperation \citep{clutton2009cooperation}, disease transmission \citep{silk2017application} or social learning \citep{hoppitt2011detecting}.

Originally, relational event models were designed for small-scale networks as their computational costs are quite extensive. In this study, we have shown how they can nonetheless be applied to more complex networks based on high-quality standardized databases of ecological data. Further developments of novel, computationally efficient implementations and user-friendly software will help quantitative ecologists to make the standard application and diffusion of the methodology possible and more widely investigate time-dependent ecological phenomena.

\section*{Supplementary Materials}

Supplementary material is available at \url{http://biostatistics.oxfordjournals.org}.

\vskip 1cm
\noindent
{\it Conflict of Interest}: None declared.

\bibliographystyle{abbrvnat}
\bibliography{references}

\newpage

\begin{table}[h!]
	\centering
	\begin{tabular}{rrr}
		Variable & Description & Data source\\ 
		\hline
		$d_{sc}(t)$ & Closest distance between regions & \citep{hijmans2017package}\\
		$tr_{sc}(t)$ & International trade & \citep{Barbieri2009}\\
		$dt_{sc}(t)$ & Near-surface air temperature & \citep{Watanabe2011}\\
		$l_c(t)$& Sum of cropland and pasture proportions  & \citep{Hurtt2011} \\
		$u_c(t)$& Proportion of urban area  & \citep{Hurtt2011} \\
		$k_{sc}(t)$ & Colonial ties  & \citep{COLDAT}\\
		$pi_{c}(t)$ & The Alien Species First Records database & \citep{Seebens2018}\\
		\hline
	\end{tabular}
	\caption{ List of variables and data sources used in analysis. Although each of the various data sources contained a wider temporal range, our analysis focused on the period from 1880 until 2005 and we subsequently focused on this temporal range for each of the databases. \label{tab:data}}
\end{table}

\begin{table}[h!]
	\centering
	\begin{tabular}{p{0.08\textwidth}p{0.28\textwidth}p{0.21\textwidth}p{0.28\textwidth}}
		Effect & Definition & Type & Justification \\
		\hline
		$d_{sc}(t)$ & The distance to the nearest region $c$ in which species $s$ is present by time $t$. & Time-varying variable with effect changing over time & Geographic distance may influence migratory behaviours and strategies\\
		$tr_{sc}(t)$ & Logarithm of the sum of annual trade flows between region $c$ and other regions invaded by species $s$ by time $t$ & Time-varying variable with effect changing over time & International trade is a key pathway for the global spread of alien species \\
		$dt_{sc}(t)$ & Minimum absolute difference in temperature between region $c$ and other regions invaded by species $s$ by time $t$ & Time-varying variable with constant effect & Climatically suitable ranges ease establishment of alien species \\
		$l_c(t)$ & Sum of cropland and pasture proportions in the region $c$ at time $t$ & Time-varying variable with constant effect & Land use changes provide opportunities for the spread and establishment of alien species\\
		$u_c(t)$ & Proportion of urban area in region $c$ at time $t$ & Time-varying variable with constant effect & Land use changes provide opportunities for the spread and establishment of alien species \\
		$k_{sc}(t)$ & Presence of species $s$ at time $t$ in a region within the area of a colonial power &  Time-varying variable with constant effect& In the Colonial era alien species have been intentionally introduced \\
		$pi_{c}(t)$ & Weighted number of prior invasions in region $c$ & Time-varying variable with constant effect & Temporal variable representing the changing importance of region $c$ over time \\
		$se_{c}$ & Sampling effort & Constant variable with constant effect & The number of native species per region is used as a proxy for sampling effort\\
		$b_s$ & Species invasiveness  & Monadic random effect & Different species may vary in their spread behaviour \\
		$b_c$ & Region invasibility & Monadic random effect &  Certain regions may be more attractive destinations than others \\
		$b_{ss'}$ & Affinity of species $s$ and $s'$ in their spread patterns & Dyadic random effect & Alien species may take similar invasion pathways \\
		\hline
	\end{tabular}
	\caption{List of explanatory variables used for model fitting. \label{tab:variables}}
\end{table}

\begin{table}[tb]
	\centering
	\begin{tabular}{rrrrrr}
		&	 & Model loglik & $\chi^2$ stat & df & p-value \\
		\hline
		\multirow{3}{*}{Birds} & Only constant effect & -16746 & && \\
		& Proposed model	& -16597  & 298.4   &  12 & $p< 10^{-4}$\\ 
		& Only time-varying effects &  -16592.6 & 8.8 &  4  &  0.07\\	
		\hline
		\multirow{3}{*}{Mammals} & Only constant effect	&-3513.6 & && \\
		& Proposed model	& -3299.9   & 427.37   &  12 & $p< 10^{-4}$\\ 
		& Only time-varying effects & -3295.7 & 8.4 &  4 & 0.08\\
		\hline
		\multirow{3}{*}{Plants} & Only constant effect & -88361.0   & && \\
		& Proposed model	& -85782.0 &  5156.7  &  12 & $p< 10^{-4}$\\ 
		& Only time-varying effects	& -85772.0 & 20.1 &  12 & 0.07\\	
		\hline
		\multirow{3}{*}{Insects} & Only constant effect & -3163.1  & && \\
		& Proposed model	& -3136.5   & 53.3 &  8 & $p< 10^{-4}$\\ 
		& Only time-varying effects	& -3132.2 & 8.5 &  8 & 0.39\\	
		\hline		
	\end{tabular}
	\caption{Summary table of the ANOVA results, showing the preference for the proposed model including several time-varying effects for all the taxonomic groups. \label{tab:comparison}}
\end{table}

\begin{table}[h!]
	\centering
	\begin{tabular}{rrrrrrrrr}
		Taxonomic group & $R^2$ & Null loglik & Model loglik & AIC & BIC & $\chi^2$ stat & df & p-value \\
		\hline
		Birds & 18.9 & -20463.1 & -16597.2 & 7695.9 & 7597.1  & 7731.9 & 18 & $p< 10^{-4}$ \\
		Mammals & 25.6 & -4437.2 & -3299.9 & 2236.6 & 2159.4 &  2274.6 & 19 &$p< 10^{-4}$ \\ 
		Plants & 26.1 & -116051.9 & -85782.3 & 60497.2 & 60348.7 & 60539.2 & 21  & $p< 10^{-4}$ \\
		Insects & 12.3 & -3577.9 & -3136.5 &  852.8 & 794.3 & 882.8  & 15 & $p< 10^{-4}$ \\ 
		\hline
	\end{tabular}
	\caption{Summary table of the final invasion models selected, showing an overall fit between 12.3\% and 26.1\% for all taxonomic groups. \label{tab:summary}}
\end{table}

\begin{table}[h!]
	\centering
	\begin{tabular}{rcccc}
		effect & plants & insects & mammals & birds \\
		\hline
		$pi_c(t)$: prior invasions & -0.005* & -0.11* & n.s. & n.s.\\
		$dt_{sc}(t)$: temperature difference  & -0.172* &-0.089* & -0.064* &-0.052* \\
		$k_{sc}(t)$: colonial ties & n.s. & n.s.  & n.s.& n.s.\\
		$u_c(t)$: urban landscape & -18.42* & n.s. & n.s. & n.s.\\ 
		$se_{c}$: sampling effort & 1.8* & 0.65* & 0.99* & n.s.\\ 
		\hline
		$\sigma_{\mbox{\scriptsize spc}}$: species invasiveness & 0.64* & 0.85* & 0.98* & 0.92* \\
		$\sigma_{\mbox{\scriptsize cnt}}$: region invasibility & 3.59* & 1.22* & 1.36* & 1.49*\\
		$\sigma_{\mbox{\scriptsize int}}$: interdependence in spread & n.s. & 0.71* & {\bf 0.22}* & 0.41* \\
		$\sigma_{\mbox{\scriptsize int}}^{\mbox{\scriptsize sym}}$: symmetric interdependence in spread &  n.s. & {\bf 0.82}* & n.s. & {\bf 0.34}*\\
		\hline
	\end{tabular}
	\caption{Parameter estimates for the non-temporal fixed effects and random effect standard deviations across the four taxonomic groups (* p-value $<0.001$, n.s.: p-value $>0.05$). The bold values indicate the preferred formulation of interdependence in spread effect in each case, i.e., the random effect with the highest trade-off between explanatory power and the number of parameters needed.}  \label{tab:est}
\end{table}

\begin{figure}[h]
	\centering
	\includegraphics[width=\linewidth]{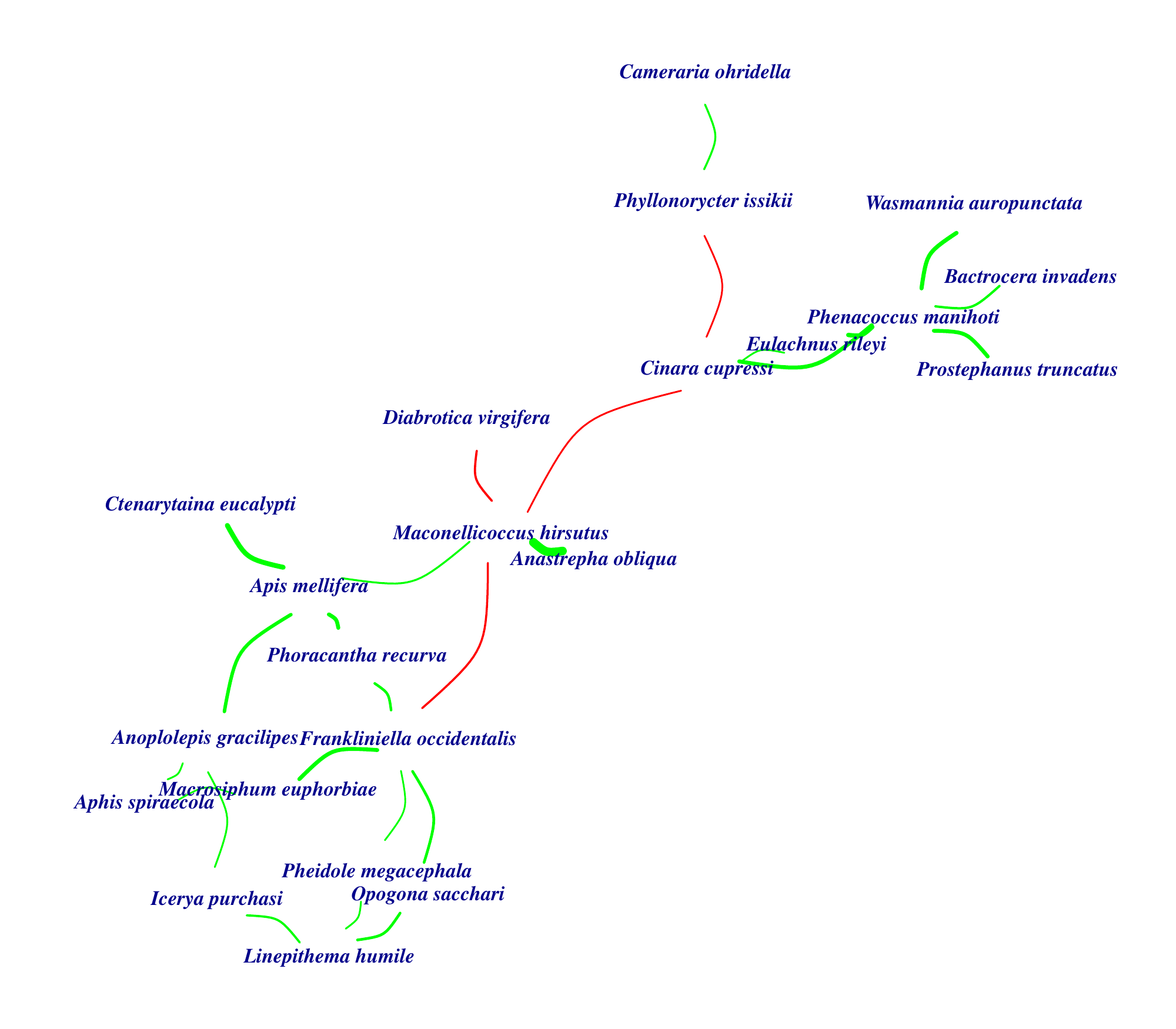}
	\caption{Symmetric insect interdependence in spread effects. Each node represents a species and an edge depicts a interdependence in spread propensity. Green edges indicate that the spread of two species tends to follow each other, while red edges suggest that two species tend to avoid each other in their spatio-temporal spread. Moreover, the thickness of the edge represents the strength of the measured interdependence. The provided graph include only the subset of edges having the highest weight values, i.e. increasing or decreasing the invasion hazard by a factor of 2. Smaller interdependence strengths have been removed to make the graph easier to interpret.}
	\label{fig:interaction}
\end{figure}

\end{document}